\pdfoutput=1

\newcommand{\degr}{$^\circ$}
\newcommand{\caproman}[1]{\uppercase\expandafter{\romannumeral#1}}
\newcommand{\bs}{$\backslash$}
\newcommand{\BGsite}{http://physics.harvard.edu/\raisebox{-4pt}{$\tilde{\;\;}$}gottschalk}
\newcommand{\BGmail}{bgottsch\,@\,fas.harvard.edu}

\documentclass{article}
\usepackage{amsmath}
\usepackage{latexsym}
\usepackage{graphicx}

\begin{document}

\title{\bf Skewness and kurtosis as measures of range mixing\\in time resolved diode dosimetry}
\author{Bernard Gottschalk
\thanks{\;\;Harvard University Laboratory for Particle Physics and Cosmology, 18 Hammond St., Cambridge, MA 01238, USA,\;\BGmail}}
\date{October 11, 2011}
\maketitle

{\centering{\Large\bf (Technical Note)}\\~\\}

\begin{abstract}
\noindent
We discuss the analysis of two time resolved diode dosimetry runs at the Burr Center. The 09MAR11 run was intended to determine $d(\sigma_t)$ for DFLR1600 diodes exposed to `scout beam' runs of $d_{80}=18$\,cm, a depth appropriate to prostate AP fields. We extract from the same data, however, functions $s(\sigma_t)$ and $k(\sigma_t)$, the `natural' dependence of the skewness and kurtosis of the pulse burst on $\sigma_t$ in the absence of range mixing. 

The 15JUN11 run studied, with a pelvic phantom, the correlation between $d(\sigma_t)$ (from scout beam measurements) and $d_{80}$ (from dose extinction measurements) in three configurations: open field, aperture plus range compensator, and solid water phantom. The tight correlations observed agree with previous findings by S. Tang even though our analysis is quite different. Discrepancies of order 1\,mm remain, hopefully to be resolved by more careful measurements in future.

However, we go further, comparing $s$ and $k$ obtained from scout beam exposures in the three configurations with their natural values. The excess $s$ and $k$, with error estimates derived from the same data, allow us to compute $p_\mathrm{ES}$ and $p_\mathrm{EK}$, the probabilities of seeing deviations that large solely from statistical fluctuations. $s$ and $k$ being independent moments, $p_\mathrm{ES}$ and $p_\mathrm{EK}$ are independent probabilities. That justifies us in taking their product $p_\mathrm{both}$ , which appears to discriminate effectively between diodes whose dose extinction results appear normal and those where range mixing is evident.

Essentially, we compare two distributions by comparing their next odd and even central moments beyond the one ($\sigma_t$) we actually use. (The $\chi^2$ test is not directly applicable because we do not know either distribution, only random samples.) Taken at face value our result means that we can, from scout data alone, and without knowing diode positions in the phantom, infer which diodes in an array see range-mixed protons and therefore report untrustworthy $\sigma_t$'s. This should, of course, be tested further in experiments where the range mixing is carefully controlled.
\end{abstract}

\vspace{5pt}
\begin{center}{\bf\small Note Added \today}\end{center}
\begin{quote}\small
This is a technical note originally intended only for my collaborators (see Acknowledgements). It describes some hardware, some software, and some data analysis including a novel way of identifying, in an array of diodes, those (if any) receiving range-mixed protons. In writing up the related experiment for publication we now require a citable reference, hence this arXiv submission. The original document is unedited except for this note. The skewness/kurtosis method has been patented (Pub. No.: US2012/0228493 A1, Pub. Date: Sep. 13, 2012). The experiment to test it in more detail has yet to be done. 
\end{quote}

\clearpage
\tableofcontents

\clearpage
\section{Introduction}\label{sec:Introduction}
This introduction is somewhat lengthy: it is designed to familiarize the general reader with the project. My collaborators may want to skip most of it.

\subsection{History}
It is a paradox of proton radiotherapy that its most distinctive feature, the sharp distal edge of the depth-dose curve (Bragg peak), is rarely used to cut between tissue to be treated and tissue to be spared. The reason is that the exact stopping point of the proton beam is usually uncertain to a few millimeters owing to such factors as uncertainty in CT number to range conversion, approximations in the dose algorithm, day to day setup, and day to day changes in internal patient anatomy. Much effort has been invested in techniques (PET imaging and prompt gamma ray detection among others) hoping to improve on this situation.

The particular idea we pursue here is not new. At the Harvard Cyclotron Laboratory around 1981, Miles Wagner inserted a dosimeter in the oral cavity of patients undergoing head-and-neck radiotherapy. If a proton beam employing a rotating range modulator is used it is possible to infer, from the shape of the dose v. time in the dosimeter, the water equivalent depth of the dosimeter. That information can be used to correct the beam energy on that day. Unfortunately, the technique (`synchroprobe') did not enjoy clinical acceptance, and was never published.

Our IBA/MGH/Harvard group is currently directing its efforts, instead, at potential AP fields in prostate radiotherapy, where the organ to be treated is the prostate and the organ to be spared is the rectal wall.\footnote{~Head-and-neck and \oe sophagus are other sites which might be considered.} The dosimeter can be attached to a rectal balloon and pressed against the rectal wall. Lu \cite{hmLuInVivo} described the technique and investigated it in a water phantom. The dosimeter (a small ionization chamber) was calibrated by recording dose v. time patterns at many depths in a water tank exposed to a `scout beam' (our term) having somewhat excess range. An unknown depth could subsequently be determined by finding which calibration pattern matched best. The technique was accurate to $\approx1$\,mm.

\begin{figure}[p]
\centering\includegraphics[width=4.55in,height=3.5in]{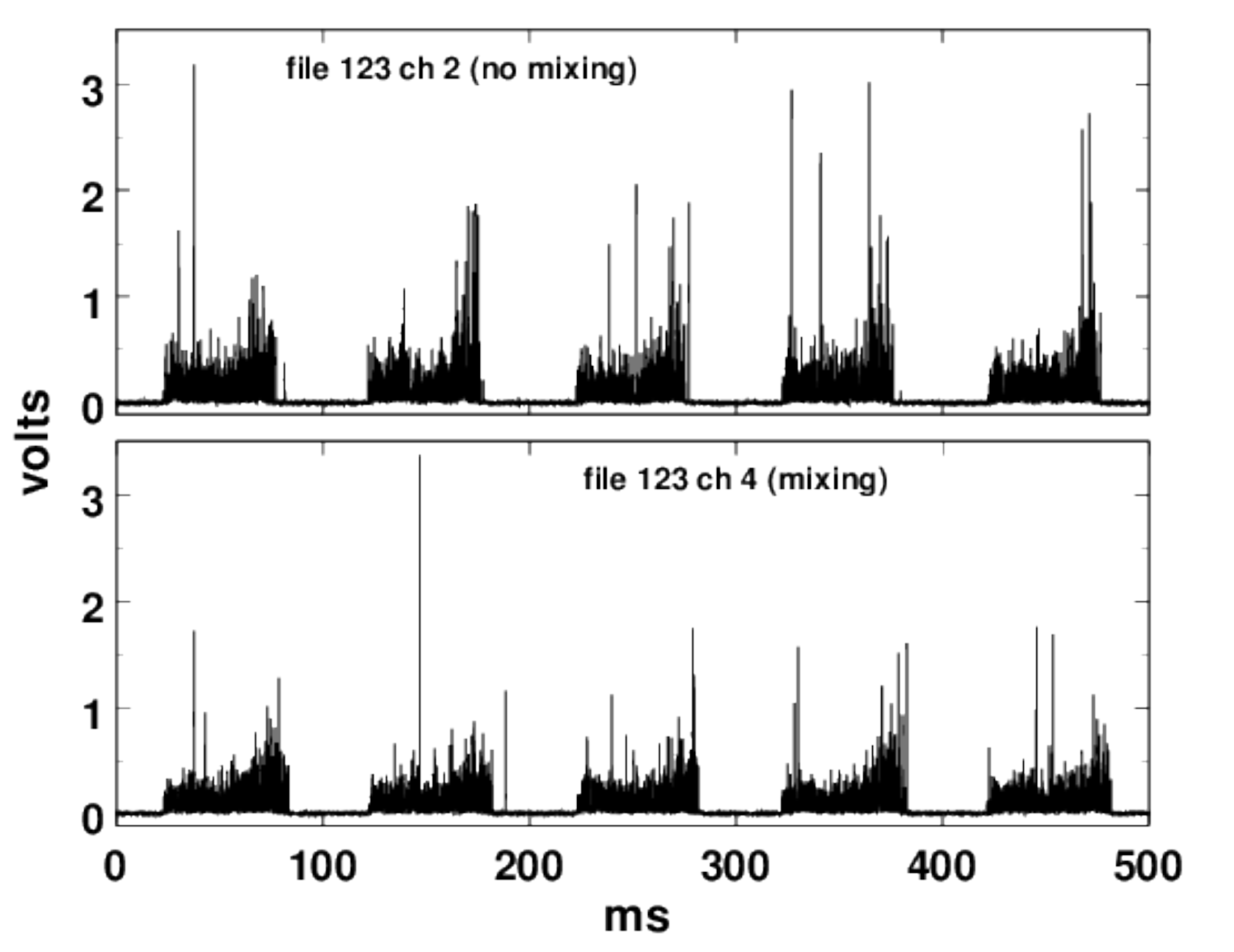} 
\caption{Top: raw data, scout beam file 123 channel 2 (AP + RC, no range mixing). Bottom: same file channel 4 (AP + RC, range mixing).\label{fig:raw123ch2ch4}}
\end{figure}

Later, our attention shifted to semiconductor diodes, being small, inexpensive, rugged two-terminal devices which do not require bias voltage. The first experimental study \cite{BGwepl} used a PTW diode designed for dosimetry, driving a reasonably fast ($\approx10\mu$\,s) current-to-voltage amplifier whose output was sampled by a digital oscilloscope. Instead of a relatively smooth dose v. time signal, a sequence of random pulses is observed (Figure\,\ref{fig:raw123ch2ch4}), each representing one or a few protons: the effective diode area is so small that the quantization of proton fluence becomes evident. Pattern matching no longer works, but it was found that a simpler method based on $\sigma_t$, the rms width of the time distribution, worked even better, yielding sub-mm accuracy. (It also works for smooth distributions as with ionization chambers.)

Since that publication, we have improved the equipment and answered a number of questions. We have built a 12-channel amplifier which drives an off-the-shelf 12-channel sampling analog-to-digital converter. That allows an array of diodes to be used rather than just one, whose position might prove unfavorable. We have found that standard, inexpensive surface-mount diodes DFLR1600 work as well as the PTW diode. Their calibration is identical, and independent of diode orientation. Being in essence a measure of the time distribution, $\sigma_t$ appears to depend on beam parameters rather than properties of the dosimeter.

The present report covers data taken earlier this year with the multi-channel equipment, first in a water tank and then, with the addition of a more efficient trigger and a beam monitor, in a pelvic phantom. The first run 09MAR11 sought to calibrate the diode array. The second 15JUN11 investigated whether the `dose extinction' (DE) method (reducing the beam range until the signal goes away) correlates with the `scout beam' (SB) method just described. We have ignored a few other runs of doubtful value.  

\subsection{Comparison With the S.\,Tang Analysis}
Shikui Tang has already discussed these data in a series of internal reports \cite{ST09MAR11,ST15JUN11,STrangeMixingPhantom,STrangeMixing} which we'll refer to collectively as ST. They cover much of the same ground. Indeed, they go further in relating what each diode observes to its position in the phantom (using X-ray and treatment planning information) whereas we will confine ourselves to what can be deduced from the diode signals alone. 

However, our methods are substantially different even where our conclusions agree. We treat the DC voltage offset problem from first principles and use it to correct the DE distributions. ST (who refers to it as `dose background') computes it by an undisclosed method, but eventually abandons that, essentially making the toe of the DE distribution zero by definition. That only works, of course, for data sets which include the toe. ST computes such quantities as $d_{80}$ by fitting a polynomial to a limited (and therefore arbitrary) region of the DE falloff, whereas we fit the entire distribution, including data from the scout beam. Thus our procedure has fewer arbitrary steps. Nevertheless, the correlations found between DE and SB measurements are very similar in the two analyses, testifying to the robustness of that result.

\subsection{Range Mixing}
The central problem of proton dosimetry---for instance, computing the dose distribution in the patient---is that, owing to the interplay of transverse heterogeneity with multiple Coulomb scattering, protons can reach the same point with different energy-loss histories and therefore, different stopping powers. In early publications e.g.\cite{Urie1986} this was called `degradation of the Bragg peak'. Nowadays it has the better name `range mixing'. It poses a serious problem to the SB method of determining the water equivalent path length (WEPL) to a given point. Even if an SB measurement yields some number for $\sigma_t$ (it always will) if there are multiple paths (WEPLs) to the diode in question that number conveys no useful information and must be disregarded. 

That is one reason we use an array rather than a single diode. The heterogeneities in a typical prostate AP plan allow us to hope that at least a few diodes will see pure rather than range-mixed protons, but we must identify those diodes. Knowing the diode positions from X-rays we might try to predict, from the treatment plan, which diodes are trustworthy, but it would be simpler operationally if we could deduce that directly from the diode signals. 

\subsection{Detecting Range Mixing Through Aberrant Moments}
The method proposed assumes only that, when range mixing occurs, it results in some change in the distribution of signal v. time for that diode. If we knew the set of distributions measured at calibration time (no range mixing) and the distribution at measurement time (suspected range mixing) we could simply compare the latter with the former using the $\chi^2$ test \cite{nr} and find, perhaps, that none of them match. But we know neither parent distribution, only random samples of each. 

However, if two distributions differ, at least one of their moments must differ, and we do have multiple samples, both at calibration time and measurement time, which yield moments and their estimated measurement errors. Accordingly, given some $\sigma_t$, we ask whether the next odd and even central moments (the skewness $s$ and kurtosis $k$) agree with their `natural' values at that $\sigma_t$, measured at calibration time. As a measure of agreement we use the quantity $p$, familiar from statistics, namely the probability of seeing a given deviation from a normal distribution purely by chance. Very small $p$ will indicate range mixing in that diode.

\subsection{Outline}

This report documents construction, experiment and analysis performed since the last publication. It also explores the skewness/kurtosis method of identifying range mixing, which looks promising but needs further work. We begin (Section\,\ref{sec:Hardware}) by describing new hardware: the home-made 12-channel amplifier, cables, and commercial 12-channel data logger. We next (Section\,\ref{sec:Software}) describe a Fortran program ArrayRMS purpose-written to analyze the data, performing calibration, dose extinction and moments analysis as necessary. It grew from a collection of {\em ad hoc} programs and, we hope, may provide the outline for a stable multi-task program in the future. Before describing the architecture of ArrayRMS, Section\,\ref{sec:Software} covers some mathematics of a general nature. Next we cover results from the 09MAR11 calibration run (Section\,\ref{sec:09MAR11}) and the 15JUN11 pelvic phantom run (Section\,\ref{sec:15JUN11}). Finally we summarize (Section\,\ref{sec:Discussion}) with some suggestions for the future. 

\section{Hardware}\label{sec:Hardware}

\subsection{12-Channel Current-to-Voltage Amplifier}
\begin{figure}[p]
\centering\includegraphics[width=3.745in,height=3.5in]{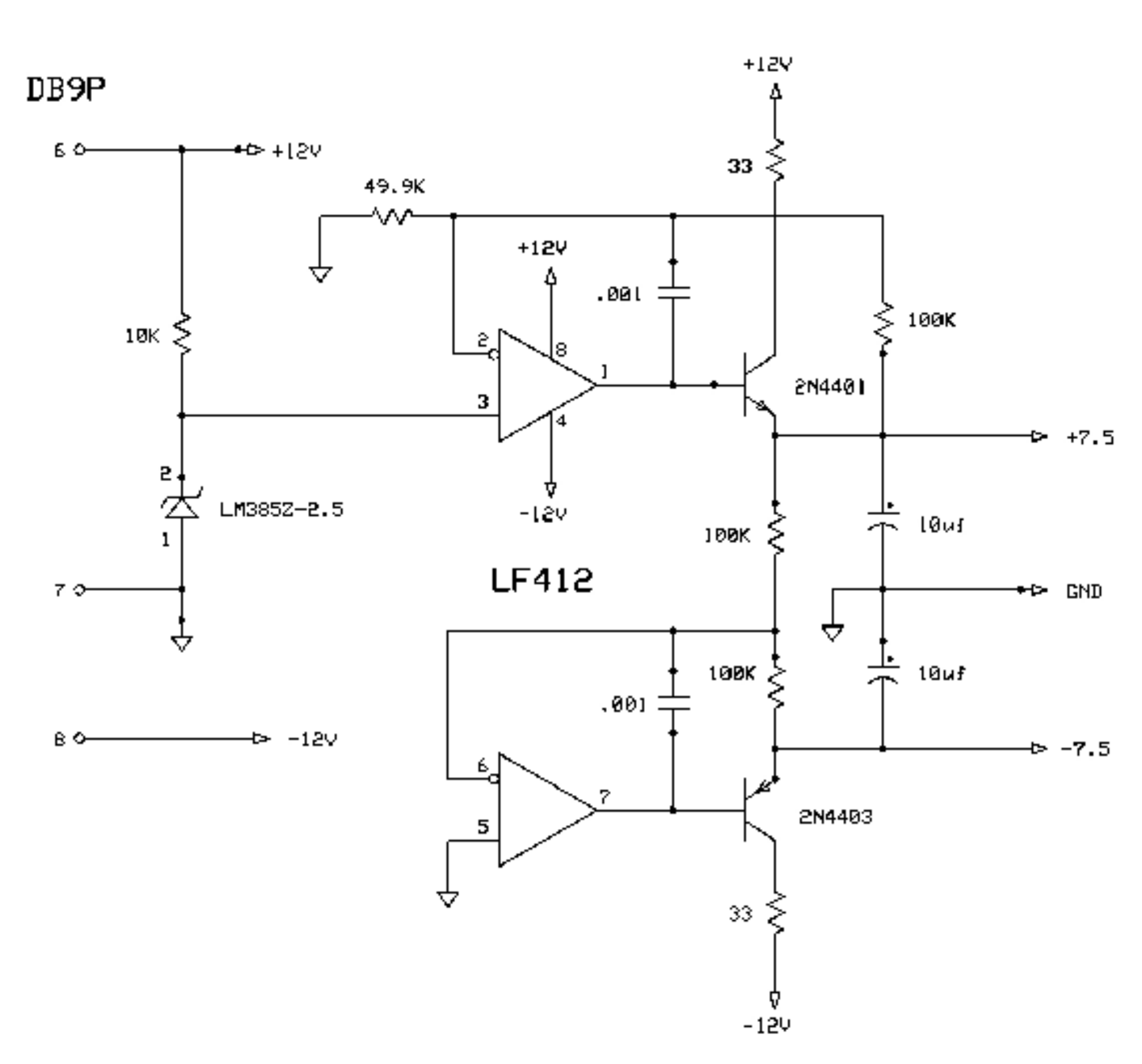} 
\caption{12-channel amplifier: on-board $\pm7.5$\,V power supplies.\label{fig:powerSupplies}}
\end{figure}

The amplifier is DC coupled and good output offset stability is important, so the 12-channel box is furnished with an on-board $\pm7.5$\,V tracking supply (Figure\,\ref{fig:powerSupplies}). It is straightforward: a precise 2.5\,V reference, a $3\times$ amplifier and an inverting amplifier. The transistors increase the available output current, while the $33\,\Omega$ resistors reduce transistor dissipation and protect the transistors against output shorts.

\begin{figure}[p]
\centering\includegraphics[width=6.3in,height=3.5in]{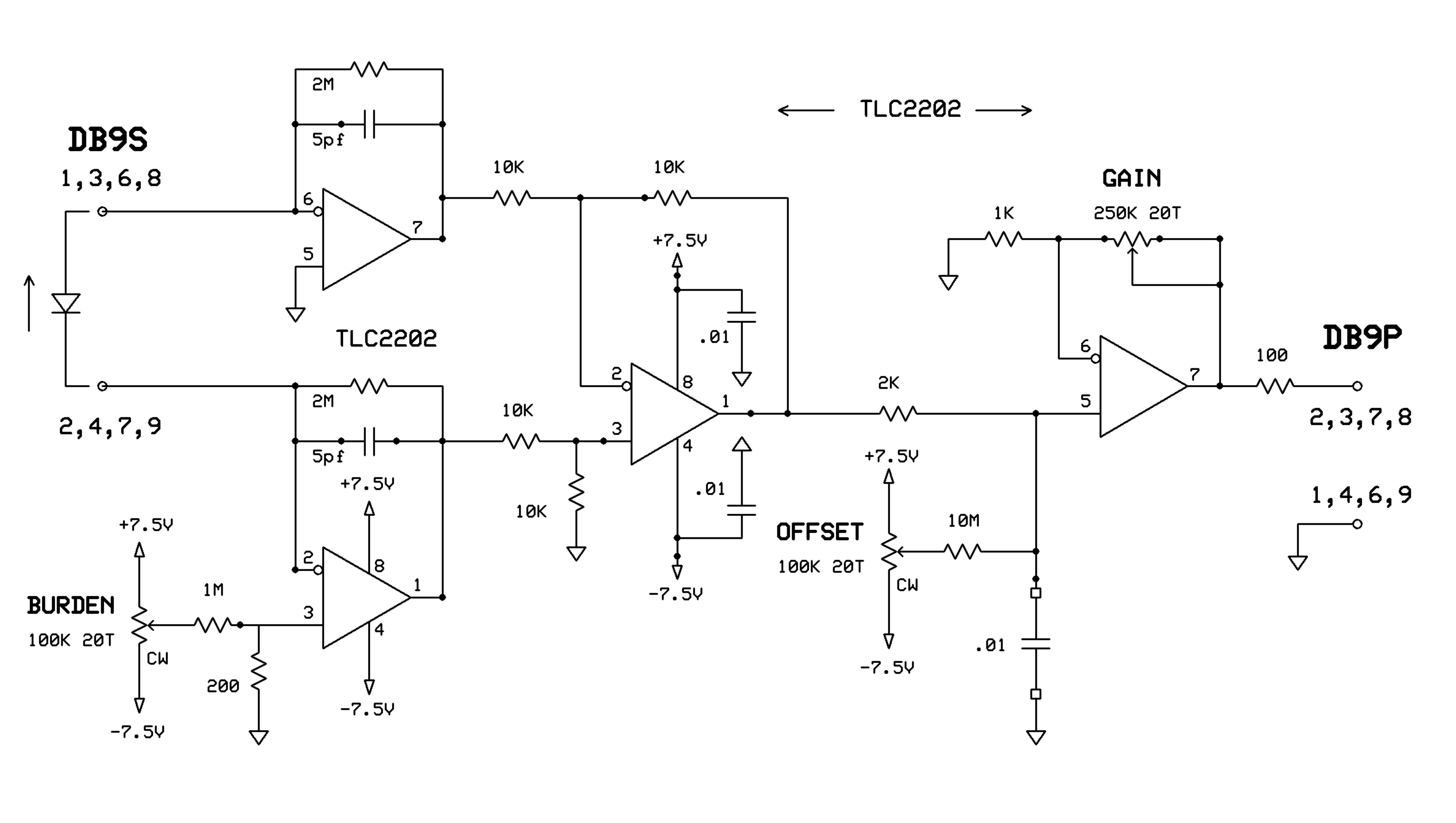} 
\caption{12-channel amplifier: one channel. DB9 pinouts are for 4 successive channels.\label{fig:12chanAmplifier}}
\end{figure}

The amplifier itself (Figure\,\ref{fig:12chanAmplifier}) consist of two transimpedance (current-to-voltage) amplifiers (gain 2\,mV/nA), a difference amplifier (gain $1\times$) and an output stage (nominal gain $125\times$). The choice of opamp is critical: we require very low bias current, low noise, low offset drift with temperature and time, and moderate speed. Corresponding TLC2202 specs (typical) are 1\,pA, 8\,nV/$\sqrt{\hbox{Hz}}$ (1\,KHz), 0.5\,$\mu$V/\degr C, 0.001\,$\mu$V/month and $f_t=1.9$\,MHz.

There are two DC adjustments. BURDEN ensures that the voltage across the diode, imposed by the amplifier inputs, is $0\,\pm\approx1\,\mu$\,V. Otherwise a `diode leakage current', actually due to the voltage burden across the finite diode resistance at $I=V=0$, will flow. BURDEN is adjusted by substituting (say) 10\,K for the diode and setting the output voltage to zero. Because a typical diode resistance is far greater, this need not be too precise.

OFFSET is adjusted by setting the output voltage to 0 with the input open. You may need to repeat BURDEN and OFFSET once or twice.

\begin{figure}[p]
\centering\includegraphics[width=5.3in,height=3.0in]{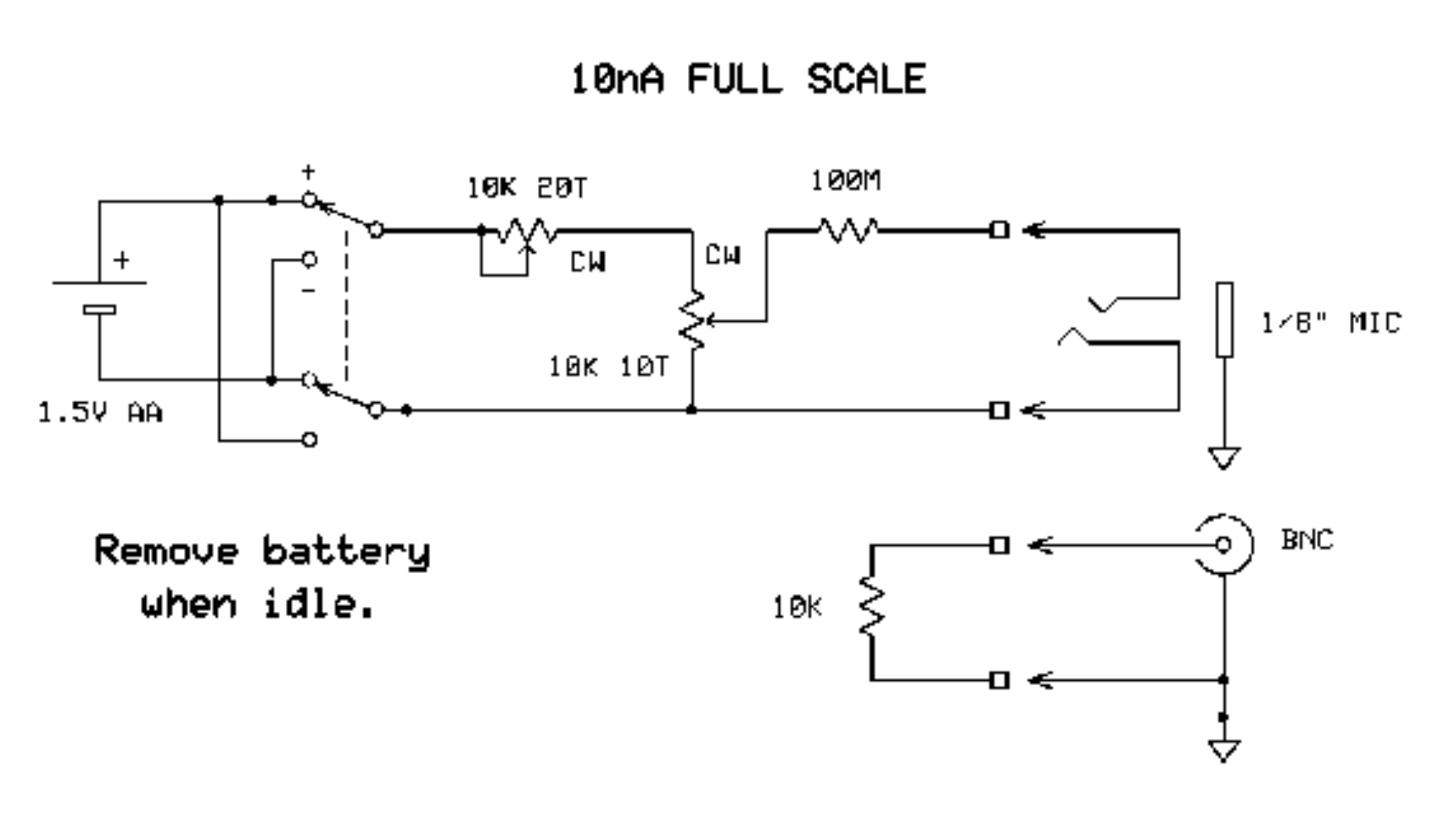} 
\caption{Calibration current source.\label{fig:calibrator}}
\end{figure}

Setting GAIN requires a well shielded, floating, balanced nA source. We built a calibration source with a battery, range setting potentiometer, front panel dial potentiometer and a large resistor inside an aluminum box (Figure\,\ref{fig:calibrator}). The switch reverses polarity to check linearity. Prior to the 15JUN11 run all channels were adjusted to 0.5\,V/nA $\pm1\%$. The precision may seem superfluous because of the large diode-to-diode variations expected, but it will allow one eventually to test for long term effects such as radiation damage to the amplifiers.

Layout is critical to prevent oscillation and crosstalk in these very high gain, fairly fast amplifiers. It goes without saying that any amplifier's output should be placed distant from its input. In addition, the 12 channels were deliberately arranged so the output of one amplifier never abuts the input of another. (The fact that the input is balanced helps, but it is never exactly balanced.) An estimate of the current induced through 0.1\,pC of unbalanced capacitance by a typical 1\,V, 20\,$\mu$s output pulse is 1V$\times\,$0.1\,pC/20\,$\mu$s\,=\,5\,nA, of the order of input currents expected. We checked for crosstalk between typical channels and none was observed.

\subsection{Cabling}
The 12-channel amplifier, placed fairly near the proton beam, is connected to the treatment control room (TCR) by twelve 75\,foot runs of
RG-178B/U miniature Teflon-insulated 50\,$\Omega$ coaxial cable, terminated in DB9S connectors (amplifier end) and male BNCs (TCR end). The cable is wound on an orange reel. In paying out or winding back the cable, one should {\em not} pass it over the side of the reel, convenient as that may seem. That introduces kinks which, if pulled tight, will eventually break the cable. One should instead grab the internal black handle and pull cable straight off the reel or, at the end of the run, wind it straight back on using the black knob. There is no effective way of repairing the cable once it is broken.

\subsection{Data Logger (IOtech\;6220)}
The data logger is an IOtech\,6220 (Measurement Computing Corp., 10 Commerce Way, Suite 1008, Norton, MA 02766, USA). We give only a brief description for background. For details, consult the Specifications and the User's Manual both of which are available at the Measurement Computing Web site.

The main feature is a 12 channel 16 bit analog to digital converter (ADC) capable of sampling each channel simultaneously at 100\,KHz. Internally, the 6220 employs three  multiplexed 4-channel ADC's so `simultaneous' actually means `within the same 10$\mu$s' but that is more than adequate. The $3\times4$ internal structure explains why channels 1--4 have a common ground and so on. The input voltage range is $\pm10$\,V so the least bit corresponds to $20/2^{16}=0.305$\,mV. The input impedance is 200\,K.

The 6220 also has flexible 8--bit digital I/O (TTL levels) brought to a rear panel DB9S connector. Lines 0--7 are connected to pins 1--8 respectively; pin 9 is ground. We built an adapter cable which makes lines 0--3 available on BNC connectors. The 15JUN11 run used line 0 as a trigger and line 1 as a counter. The 09MAR11 run used neither. 

Line 0 was connected to the facility `BOXB' signal (essentially a `beam on' signal) via a level translator furnished by Ethan. Although the software trigger settings were not changed during the run, the time at which beam actually appears, as evidenced by diode pulses, varied from one 20--burst (2\,s) sample to the next. As a result the {\em absolute} timing between pulses and BOXB, which might conceivably be useful, appears to be of no value in this run and was not used. This variable trigger delay is not understood at present.
  
Line 1 was driven by the same signal that drives the DCEU input via an interface supplied by Ethan. It therefore served as a beam monitor with 300 counts = 1 monitor unit (MU) as presently used at the Burr Center. When a line is set up as a counter, it evidently drives an internal scaler whose output is sampled at the global rate, here 100\,KHz. Since a 20 bunch (2\,s) sample only corresponds to a few hundred counts, the text file might be expected to contain many instances of 0 followed by many instances of 1 and so on: a staircase starting at 0. However, we did not generally observe 0 as the start count but instead, some small number varying from one 2\,s sample to the next. This, too, is not yet understood. We assumed that the {\em difference} between the last and the first step reflected the true number of DCEU counts.

\subsection{Data Logging Software}
Data logging software called Encore\,1.1 is supplied with the IOtech\,6220. It is launched by double clicking the Encore project icon on the Toshiba R705 desktop. The project can be edited and re-saved to preserve settings. There is reasonably good online help, and we will not attempt to describe the software in detail.

As set up for the last data run (15JUN11) described here, we took 200\,K samples (2\,s) on each of 12 channels at 100\,Ksamples/sec. One needs to turn off the simultaneous display to do this; otherwise buffer overflow will occur. Data acquisition was triggered by the BOXB signal (digital I/O line 0) and the DCEU signal was counted (totalized) as a beam monitor (line 1). After acquisition, text files were exported automatically. (Encore initially logs data in a proprietary format which is not generally useful.)

For setup runs, data logging (as opposed to display) can be toggled on/off in the main menu. The distinction between single shot operation and ordinary start/auto stop is not totally clear, and one is well advised to practice a few times and check the logged files before commencing serious work. Encore will compute and display waveform parameters such as mean and rms voltage, but the machinery is somewhat cumbersome and we did not use it except on the bench. Displaying the logged waveform in playback mode from time to time proved sufficient to monitor data integrity. Eventually, pre-analysis by ArrayRMS should be available, for the same purpose.

\section{Analysis Software (ArrayRMS.FOR)}\label{sec:Software}
Fortran program ArrayRMS.FOR analyzes the data. At present it is somewhat cluttered because of frequent changes in hardware configuration, data storage and analysis goals during this early phase of the research program. However, the basic structure seems sound. As things settle down it will be possible to trim away some of the complications. The current version of ArrayRMS, along with all supporting program modules, can be found in
 \begin{center}\bs BGware\bs source \quad at\qquad\BGsite\end{center}

The Toshiba\,R705 notebook used for data collection runs 64 bit Windows\,7 and will not run our 32 bit Fortran compiler. It will, however, run the executables produced. Our long term plan is to copy ArrayRMS.EXE to the R705, allowing one to preprocess the (very long) data files without transferring them to another computer. This step generates summary TXT files which have all the intermediate results necessary for further analysis (see Section\,\ref{sec:prelim}), and will also allow one to see how a data-taking run is progressing. However, all analysis for this report was done on our personal computer, a Lenovo\,T400 notebook running Windows\,XP Professional. Either machine processes one raw data file (2\,s worth of data or 20 bursts or $\approx2.5$\,Msamples) in less than 1\,s. Data from a three-hour run takes a couple of minutes to pre-process (task PRELIM described below).

This section begins with some mathematical considerations and closes with a short description of ArrayRMS from the user's point of view.

\subsection{Baseline DC Voltage Correction}\label{sec:vcorr}
The current-to-voltage amplifier must be DC coupled. Otherwise, rate dependent baseline shifts would make data analysis impossible. Given the high amplifier gain, it is inevitable that DC output voltages of order $\pm1$\,mV will, after some time, appear on some channels even if initially zeroed, as was done. Assuming $1$\,mV, an amplifier gain of $0.5$\,V/nA, and an integration time of 2\,s (20 bursts) the associated spurious charge $=\int i\,dt\approx\Delta t\,\sum_j i_j$ is
\[1\,\hbox{mV}\times\frac{1\,\hbox{V}}{1000\,\hbox{mV}}\times\frac{1\,\hbox{nA}}{0.5\,\hbox{V}}\times
  \frac{1000\,\hbox{pA}}{1\,\hbox{nA}}\times2\,s\;=\;4\,\hbox{pC}\]
Since it will turn out that the total charge collected in 20 bursts is of order 100\,pC and the baseline DC voltage can be several mV, this correction is small but not negligible. Its most obvious effect is that, in dose extinction measurements, the charge detected does not approach zero as the beam range decreases. The effect on scout beam measurements (that is, on $\sigma_t$) is less clear, and may be very small.\footnote{~ST calls this the `dose background' correction, but it has nothing to do with dose {\em per se}. His procedure for finding it is not described in any of the memos. Eventually ST abandons it, essentially {\em defining} the level to which the dose approaches as zero. Though that seems all right in the sense that ST's eventual results agree rather well with ours, our method seems more defensible (less arbitrary) and, of course, allows one to analyze the data properly even if measurements do not include the toe of the distribution.}

\begin{figure}[p]
\centering\includegraphics[width=4.55in,height=3.5in]{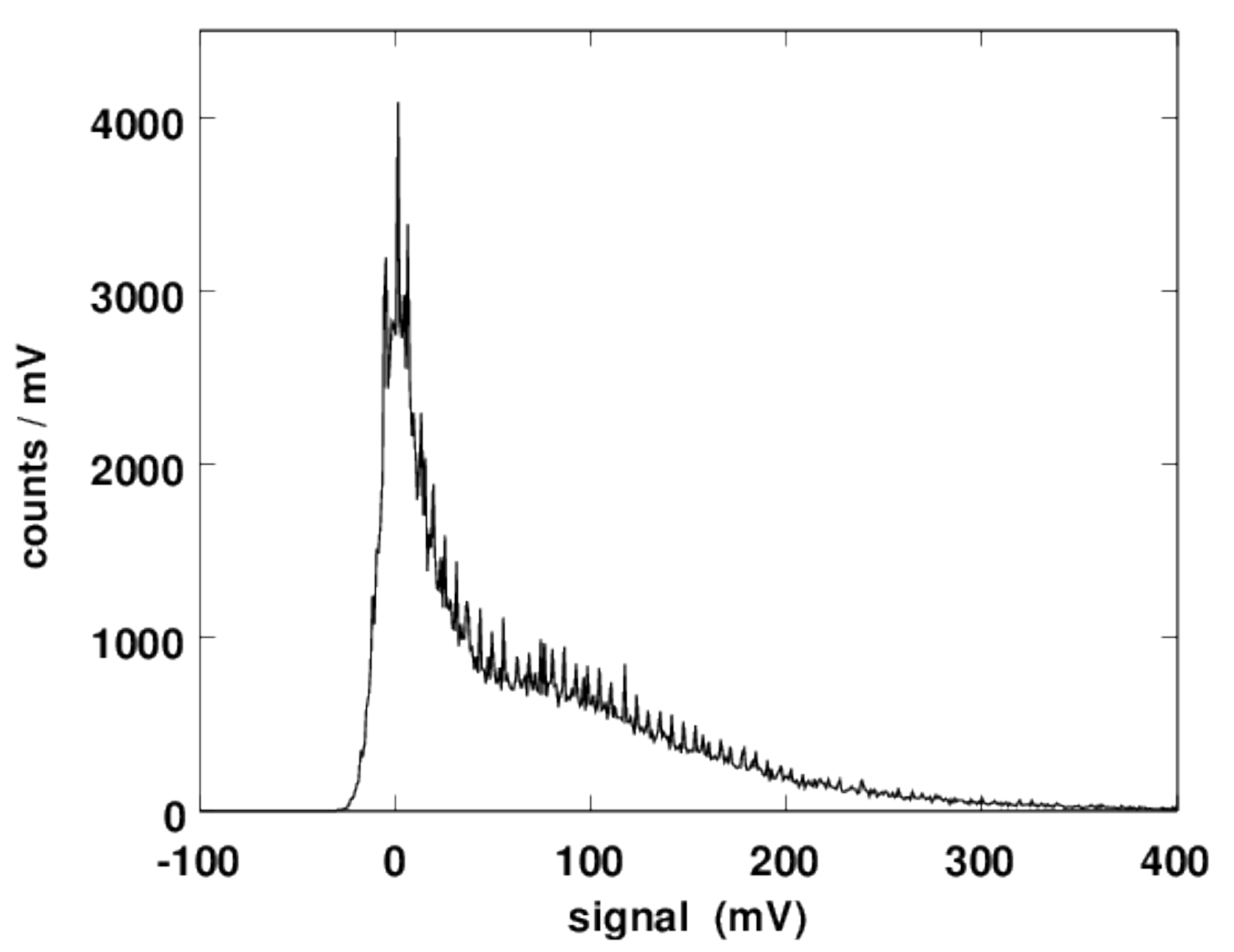} 
\caption{Unsmoothed signal spectrum with 1\,mV binning.\label{fig:freq052_1_3_0_90}}
\end{figure}

The correction is named vcorr (here $v_\mathrm{corr}$) in our software. It will eventually be subtracted from each voltage reading before any further analysis. $v_\mathrm{corr}$ is found channel-by-channel by analyzing the frequency distribution (spectrum) of all 200\,K voltage readings (assuming 20 bursts) in three successive passes. The first pass (not shown) serves merely to establish a rough value, typically very near zero. The second pass zooms in on the spectrum, using a much smaller bin width. Figure\,\ref{fig:freq052_1_3_0_90}, with a bin width of $1\,$mV, illustrates one pitfall. The least bit of the ADC equals $20/2^{16}=0.305\,$mV and any bin width of that order will yield spikes (Figure\,\ref{fig:freq052_1_3_0_90}) owing to the differential nonlinearity which is characteristic of successive approximation ADC's. 

\begin{figure}[p]
\centering\includegraphics[width=4.55in,height=3.5in]{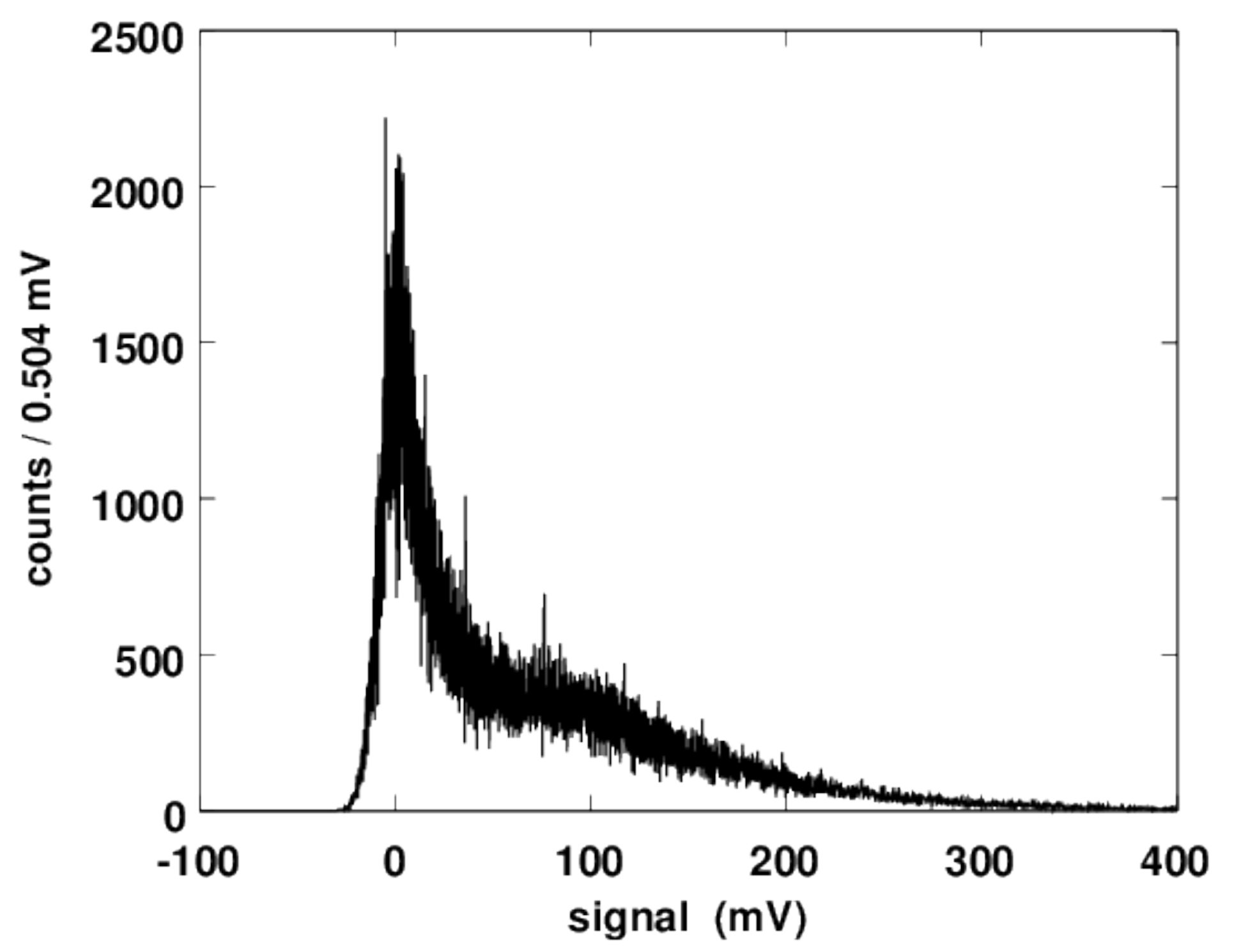} 
\caption{Like previous figure, but 0.504\,mV binning.\label{fig:freq052_504_3_0_90}}
\end{figure}

\begin{figure}[p]
\centering\includegraphics[width=4.55in,height=3.5in]{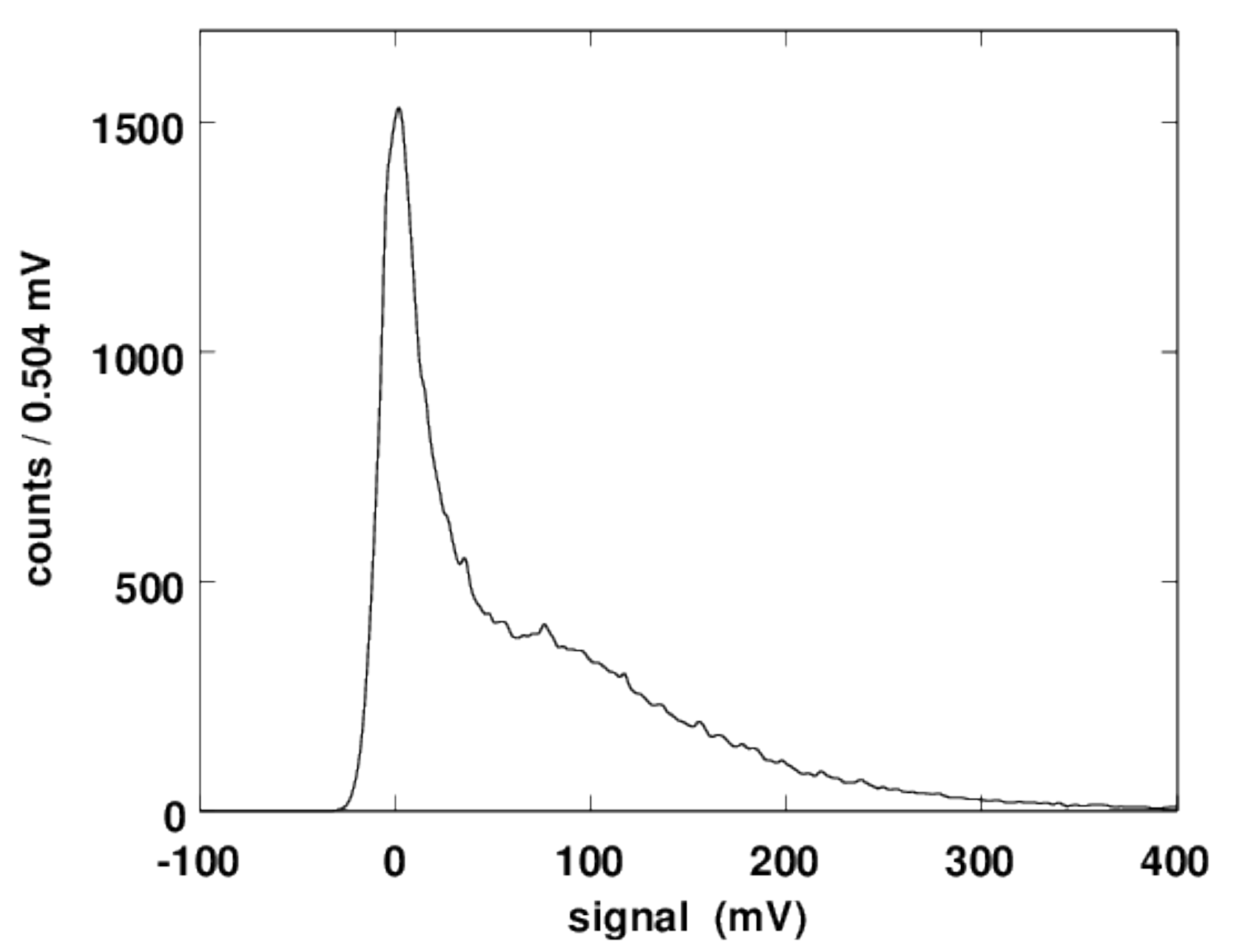} 
\caption{Like previous figure but smoothed by folding with a Gaussian of $\sigma=3$\,channels.\label{fig:freq052_504_3_1_90}}
\end{figure}

We will need to smooth the spectrum, and the spikes make it difficult to smooth. We investigated the quantization by looking at the distribution of differences between successive voltage readings, and found that $0.305\,$V and $0.336\,$V came up frequently. We do not understand the latter value, but using a bin width of $1.5\times0.336=0.504\,$V consistently gave a spectrum like Figure\,\ref{fig:freq052_504_3_0_90}, converting the spikes to an odd/even pattern which is easy to smooth (Figure\,\ref{fig:freq052_504_3_1_90}). 

\begin{figure}[p]
\centering\includegraphics[width=4.55in,height=3.5in]{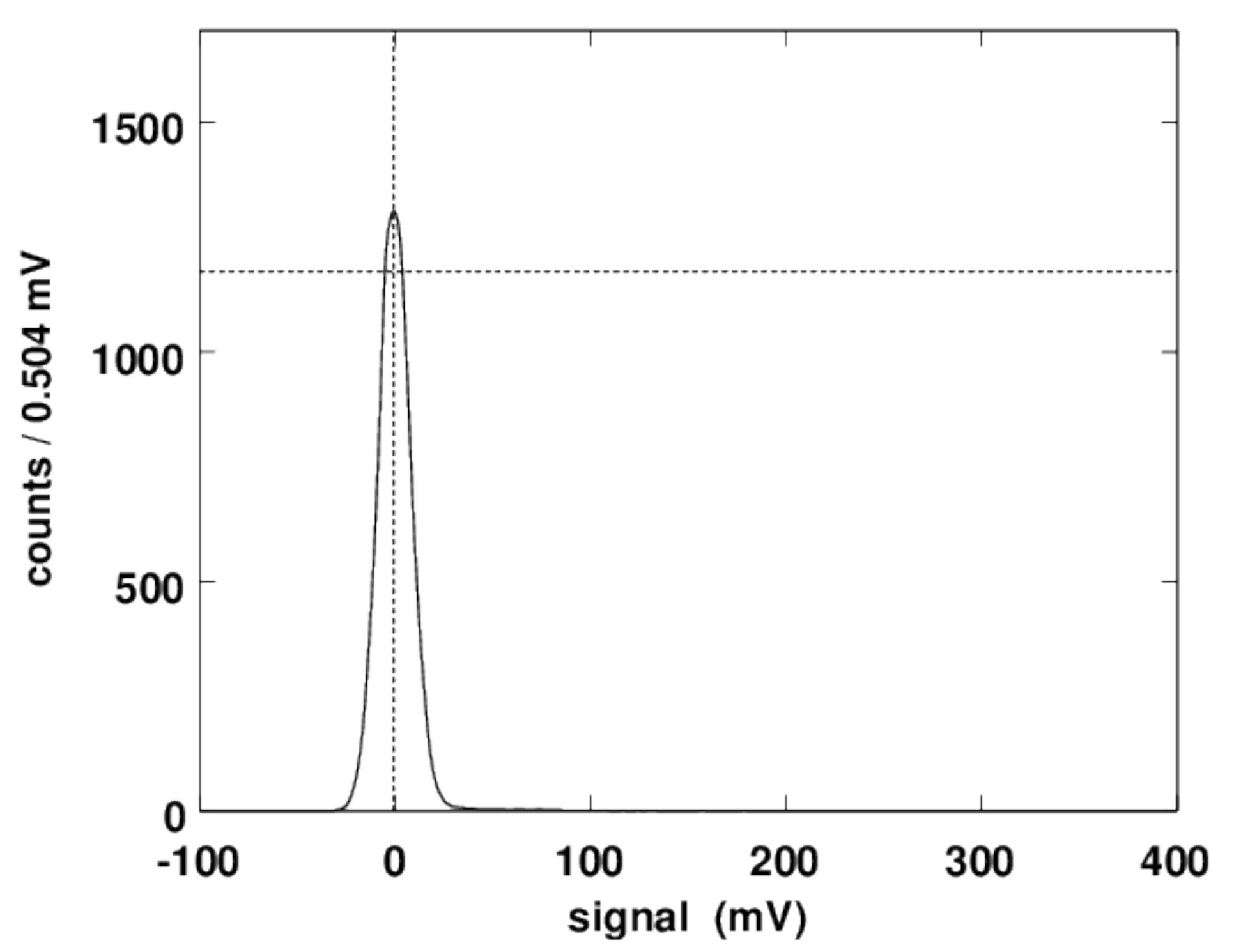} 
\caption{Like previous figure without the contribution from bursts. The mean is computed for that part of the spectrum above 90\% of the maximum.\label{fig:freq053_504_3_1_90}}
\end{figure}

However, Figure\,\ref{fig:freq052_504_3_1_90} shows a positive tail, due to signal pulses, which will have a non-negligible effect on the mean of the distribution even if we compute it for a narrow region about the peak. We therefore proceed by looking for bursts. If we find the expected 20 (or whatever) we exclude the in-burst signal, recompute the spectrum, and compute its mean over a fairly narrow region (Figure\,\ref{fig:freq053_504_3_1_90}).

\begin{figure}[p]
\centering\includegraphics[width=4.55in,height=3.5in]{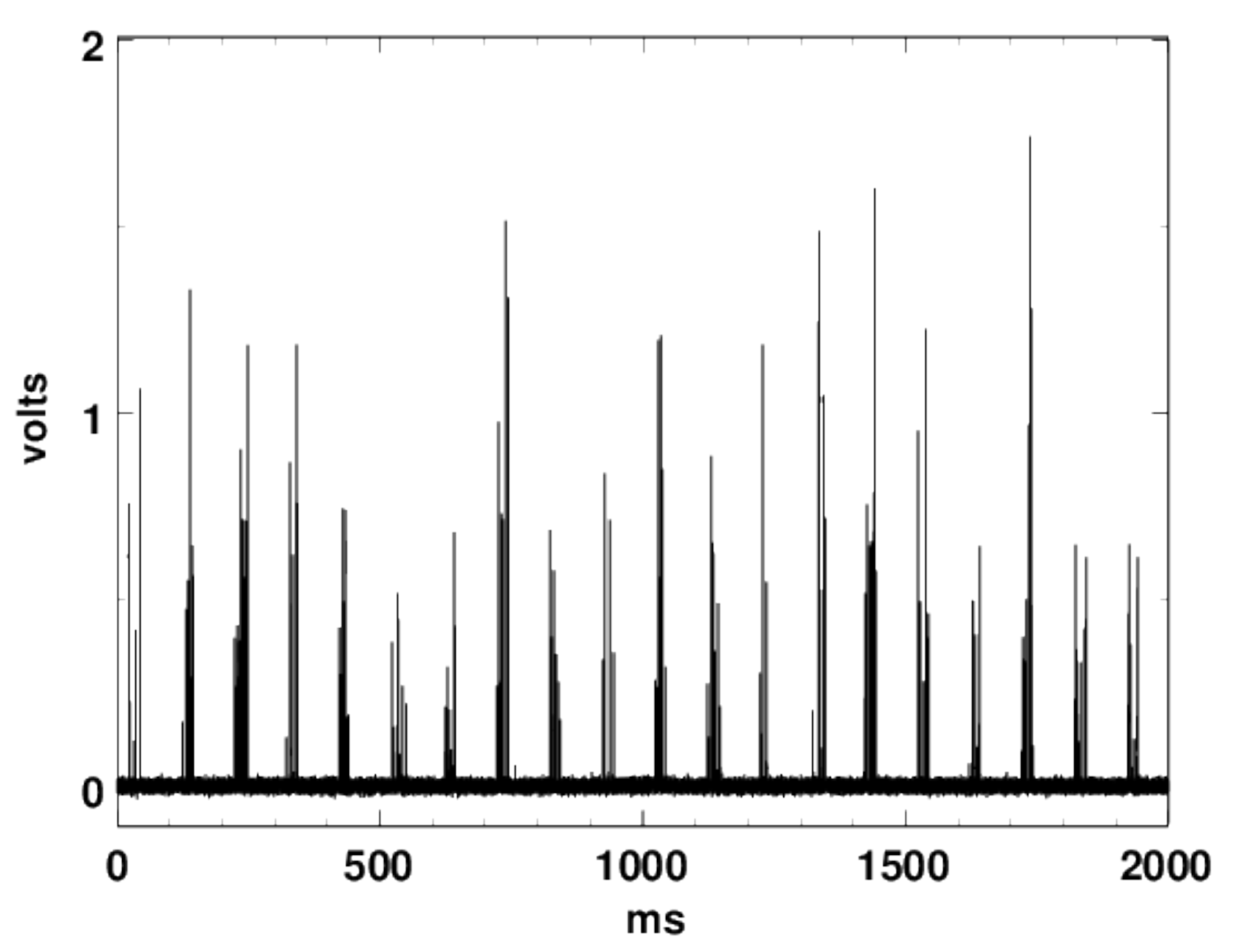} 
\caption{Raw data for the first point where the program fails to detect 20 bursts in a dose extinction series (run\,093340, channel\,7).\label{fig:rawData_55_ch7}}
\end{figure}

\begin{figure}[p]
\centering\includegraphics[width=4.55in,height=3.5in]{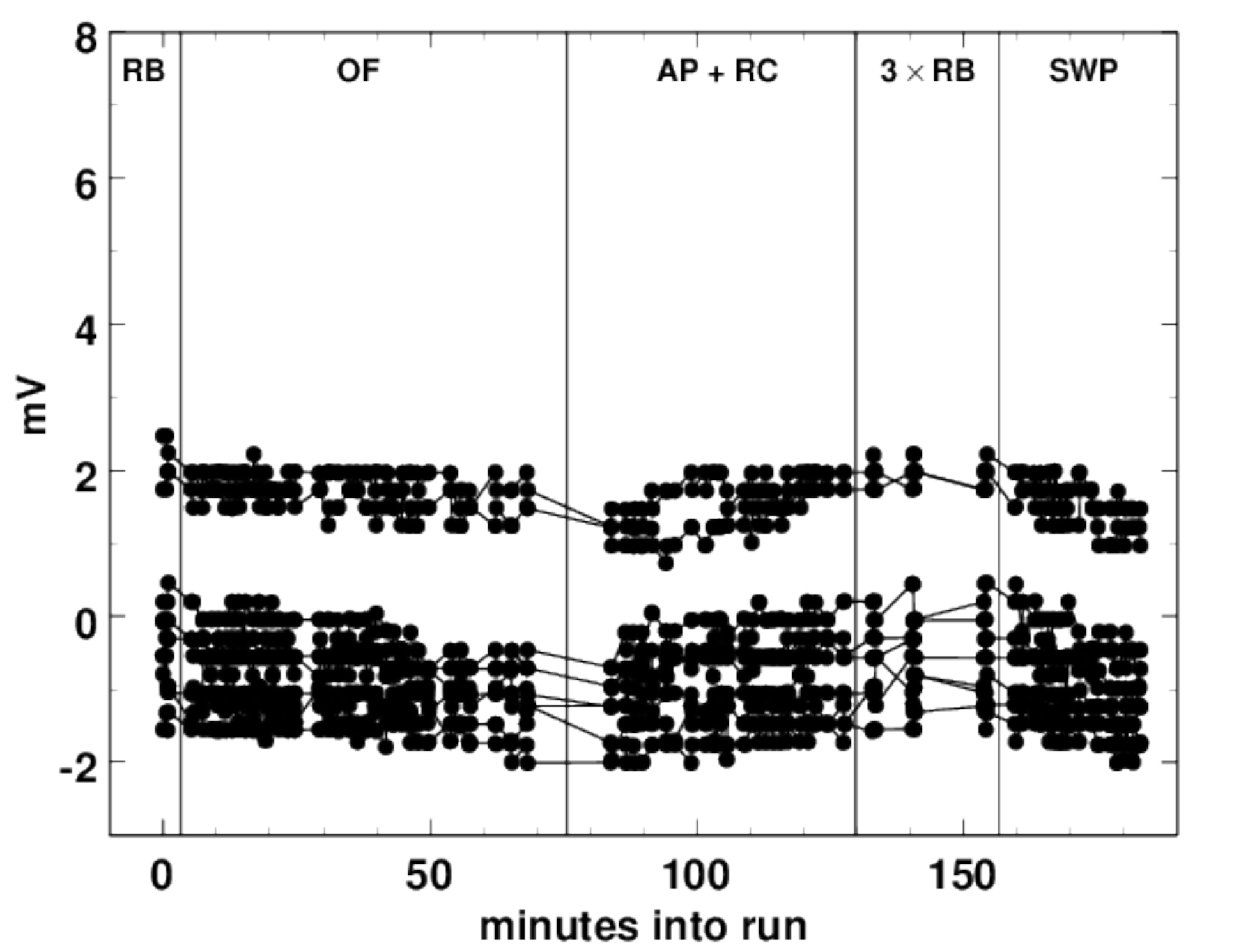} 
\caption{Evolution of $v_\mathrm{corr}$ during the run for the eight functioning channels, when in-burst signal is excluded.\label{fig:vcorr3pass}}
\end{figure}

\begin{figure}[p]
\centering\includegraphics[width=4.55in,height=3.5in]{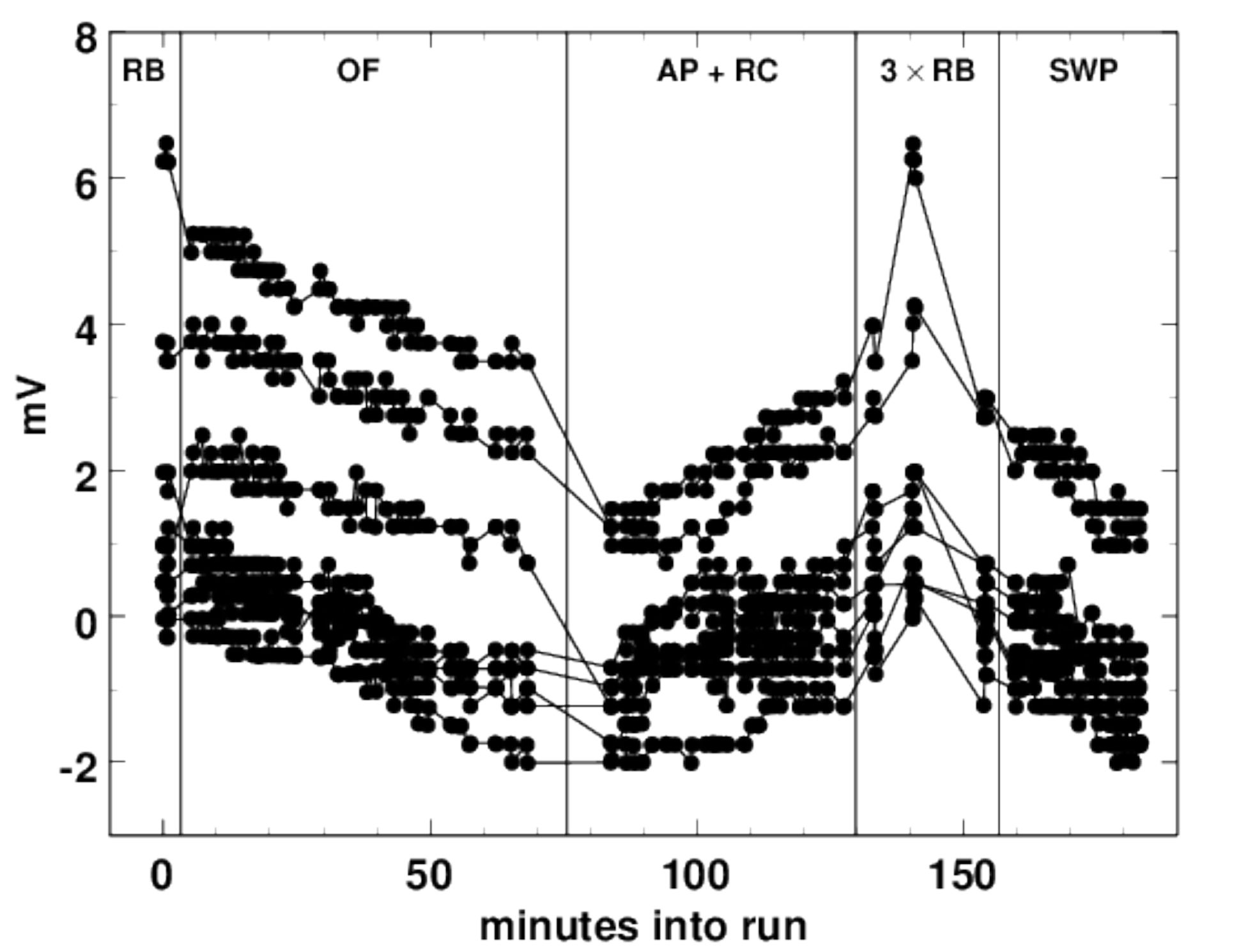} 
\caption{Evolution of $v_\mathrm{corr}$ during the run for the eight functioning channels, when in-burst signal is included in the spectrum. The dependence of $v_\mathrm{corr}$ on run conditions is spurious.\label{fig:vcorr2pass}}
\end{figure}

There is a borderline subset of data where bursts are still present but the burst finding algorithm fails to detect the expected number. Here it can probably be argued that a sufficient fraction of the trace is baseline (Figure\,\ref{fig:rawData_55_ch7}) that burst removal is not so important. Indeed Figure\,\ref{fig:vcorr3pass}, a history of $v_\mathrm{corr}$ through the 15JUN11 run, shows very little correlation of $v_\mathrm{corr}$ with the measurements being made, in contrast with Figure\,\ref{fig:vcorr2pass}, the same history with the in-burst signal left in. The dependence of $v_\mathrm{corr}$ on run conditions is spurious.

\subsection{Moments of a Distribution}
In a notation suitable to our application, the conventional definitions of the first five moments of a distribution are \cite{nr}
\begin{eqnarray}
\hbox{sum}\;\equiv\;S&\equiv&\sum_{i=i_1}^{i_N}v_i\qquad\mathrm{(V)}\label{eqn:sum}\\
\hbox{mean}\;\equiv\;m&\equiv&\frac{1}{S}\;\sum_{i=i_1}^{i_N}v_i\,t_i\qquad\mathrm{(ms)}\\
\hbox{rms deviation}\;\equiv\;\sigma_t&\equiv&\left(\frac{1}{S}\;\sum_{i=i_1}^{i_N}v_i\,(t_i-m)^2\right)^{1/2}\qquad\mathrm{(ms)}\\
\hbox{skewness}\;\equiv\;s&\equiv&\frac{1}{S\;\sigma_t^3}\;\sum_{i=i_1}^{i_N}v_i\,(t_i-m)^3\qquad\mathrm{(dimensionless)}\\
\hbox{kurtosis}\;\equiv\;k&\equiv&\frac{1}{S\;\sigma_t^4}\;\sum_{i=i_1}^{i_N}v_i\,(t_i-m)^4\;-\;3\qquad\mathrm{(dimensionless)} \label{eqn:kurtosis}
\end{eqnarray}
$\sigma_t$, $s$ and $k$ as defined here are `central' moments: they are computed relative to the mean. $v_i$ is the corrected voltage ($v_i=\hbox{reading}-v_\mathrm{corr}$) sampled at time $t_i$ and $i_1,\,i_N$ are the first and last indices of the burst. If the number of bursts detected equals the number expected (e.g. 20), PRELIM computes each statistic separately for each burst, finally returning the mean of those values and their rms deviation from the mean, a measure of the random error in that statistic for that channel.

\subsection{Combining Repeated Measurements}
Eqs.\;\ref{eqn:sum}--\ref{eqn:kurtosis} define how a given moment is computed for a single burst. It remains to describe how ArrayRMS computes the moments and estimates their errors for repeated measurements. Let us, by way of example, focus on a single channel and consider skewness and its rms deviation from the mean ($s$ and $\sigma_s$) as measured in $M$ repeated files each comprising $N$ bursts. On 15JUN11 we took, for each SB measurement, four files ($M=4$) of 20 bursts each ($N=20$) for $M\times N=80$ measurements in all. 

Initially, statistics are computed burst by burst. Next, rather than averaging all 80 measurements, that is, computing 
\[<s>_{80}\hbox{\quad and\quad}\sigma_{s,\,80}\]
it proves more convenient to proceed file by file, averaging over 20 bursts to obtain 
\[<s>_{20}\hbox{\quad and\quad}\sigma_{s,\,20}\] 
for each. Then, in order to use all the data, we average over four files to obtain
\[<<s>_{20}>_4\hbox{\quad and\quad}<\sigma_{s,\,20}>_4\]
For $\sigma_t$, $s$ and $k$ themselves this procedure makes no difference as it is easily shown that  e.g
\[<<s>_{N}>_M\;\equiv\;<s>_{M\times N}\] 
However, except for the trivial case $M=1$,
\[<\sigma_{s,\,N}>_M\;\;\ne\;\;\sigma_{s,\,M\times N}\]

\begin{figure}[p]
\centering\includegraphics[width=4.55in,height=3.5in]{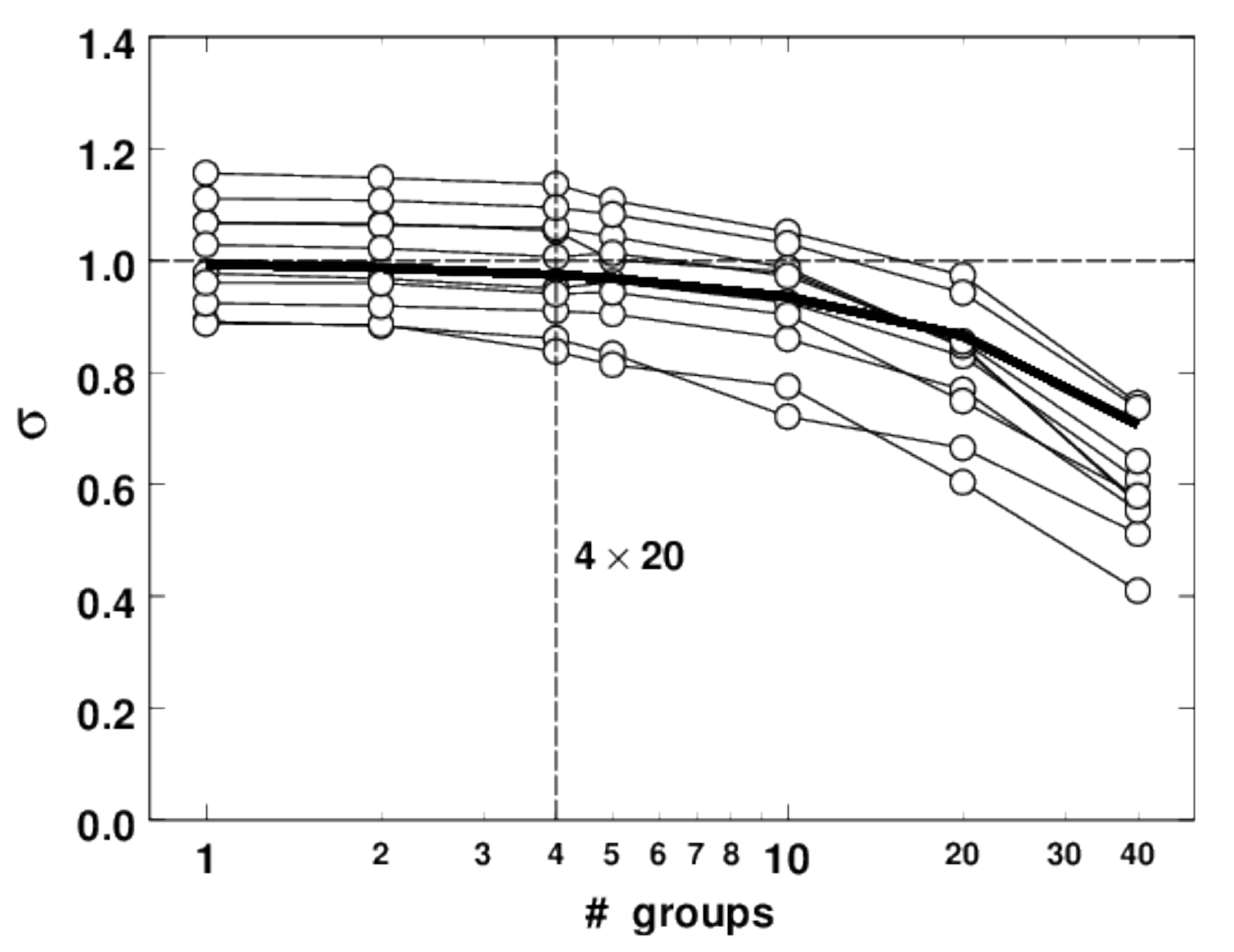} 
\caption{Effect, on the computation of $\sigma$, of dividing the data into groups. Eighty numbers were chosen randomly from a parent Gaussian having $x_0=0$ and $\sigma=1$. The experiment was repeated ten times (open circles). Bold line: plot of $\sqrt{(N-1)/N}$.\label{fig:groupRMS}}
\end{figure}

To investigate, we picked 80 numbers at random from a Gaussian of mean $x_0=0$ and rms deviation $\sigma=1$, analyzed them in $M=1,2,4,5,\,\ldots\,,40$ groups of $N=80,40,\,\ldots\,,2$\,, and repeated the experiment ten times. As Figure\,\ref{fig:groupRMS} shows, $<\sigma_{s,\,N}>_M$ decreases with $N$, simply because $\sigma$ for small samples underestimates the parent $\sigma$ by the well known `bias' factor $\sqrt{(N-1)/N}$, shown for comparison \cite{bevington}. (Evidently it is only approximate.) For $N=20$ this correction is 2.5\%, and we ignore it. 

\subsection{Computation of $p$}\label{sec:p}
Our hypothesis is that, in the presence of range mixing, the skewness of the burst distribution at a given observed $\sigma_t$ will differ from its `natural' value, that is, from the function $s(\sigma_t)$ observed in a homogeneous water phantom; also, that the same holds true for kurtosis; and finally, that skewness and kurtosis are independent. Put another way (and focusing on skewness for now), in the absence of range mixing the mean excess skewness will be zero, with some statistical error that we are in a position to estimate. If in a given case the excess skewness deviates from zero, the statistic $p_s$ is the probability of a deviation that large occurring at random. The smaller $p_s$ is, the greater the likelihood that something other than random fluctuation is going on. 

To compute $p_s$ we need the excess skewness itself and an estimate of the error in the mean of the skewness distribution. We know the rms spread in the distribution of $s$ namely $<\sigma_{s,\,20}>_4$ (with a negligible correction factor). Let us guess that the rms error in the natural function $s(\sigma_t)$ is comparable to  $<\sigma_{s,\,20}>_4$, introducing a factor $\sqrt{2}$. The error in the mean of a distribution is $\sigma/\sqrt{N}$ where $N$ is the number of measurements in the distibution \cite{bevington}. Putting all that together
\begin{equation}
\hbox{error in mean}\;=\;\sqrt{2}\times\frac{<\sigma_{s,\,20}>_4}{\sqrt{80}}
\end{equation}
The $p$ value is
\begin{equation}
p(x)\;=\;1\;-\;\frac{1}{\sqrt{2\pi}}\int_{-x}^x\,e^{\displaystyle{-\frac{1}{2}\,t^2}}\;dt\;=\;
  1\;-\;\mathrm{erf}\,\left(\frac{x}{\sqrt{2}}\right)
\end{equation}
where $x$ is the observed deviation divided by the error in the mean.

For a concrete example we look ahead to Table\,\ref{tbl:scoutResults}, the entries for channel 4 of the AP+RC configuration. The observed excess skewness is 2.4\% and the error in the mean is 0.961\% giving $x=2.51$, a $2.51\sigma$ effect with $p\,(2.51)=1.2\%$ as listed in the column labeled $p_\mathrm{ES}$. Together with an even smaller $p_\mathrm{EK}$ from excess kurtosis this yields a product $p_\mathrm{both}=0.00003\%$, ruling out a mere statistical fluctuation.  

\subsection{Fitting Procedures}

\subsubsection{Polynomial}\label{sec:poly}
The diode calibration function $d(\sigma_t)$ (or WEPL(t$_\mathrm{RMS}$) or simply WEPL) is a five term (fourth degree) polynomial. Figure\,\ref{fig:WEPLparms} (top) shows a sample parameter file. Line 1 is the number of terms and line 2 gives two parameters (midpoint of $x$ range and $du/dx$) used in the linear transformation of $x$ described next. Line 3 is the array of coefficients and line 4 is an array of corresponding error estimates \cite{bevington}. Figure\,\ref{fig:WEPLparms} also shows code to read back the fit parameters, code to evaluate the fit, and the equivalent algebraic formulas.

Instead of directly fitting a polynomial to $y_i(x_i)$ we first transform $x$ to a linearly related variable $u$ such that $-1\le u\le1$. That makes the polynomial coefficients easier to understand: each is simply the size of that term in the fit at the positive endpoint $u=1$, and its relative importance can be seen directly. (For instance, the fifth term (both from its magnitude and its error estimate) is not really necessary here, but it does no harm.) More important, the transformation renders the matrix to be inverted during the computation much better conditioned, permitting high-order fits which would otherwise fail. (An even better linear transformation, in that regard, is one which makes the standard deviation of the transformed $x$ variable unity.) The transformation is not strictly necessary here, but we have made it standard practice for polynomial fits.

\subsubsection{Broken Spline}
We handle functions not suitable for polynomials (dose extinction curves and natural skewness and kurtosis) by means of cubic spline or broken spline fits. A very brief description follows. For a full one see FitDD.pdf in \bs BGdocs at \BGsite\;.

To perform a cubic spline fit to a set of data points $y_i(x_i)$ (which, for our routines, must have monotonically increasing $x_i$ with no duplicate values) pick a much smaller set of `spline points' which are similar to instances of the data set, and whose coordinates will be the adjustable parameters of the fit. Adjust those so a cubic spline passed through them approximates the data set in the usual (least squares) sense. Interior spline points can move vertically and sideways, but end spline points can only move vertically: otherwise the fit will collapse.

The hardest (and most application specific) part of the procedure is picking reasonable initial spline points. See FitDD.pdf for details.

If the data to be fit include corners, a single cubic spline (which is smooth---no corners---by definition) will not do. In that case, fit with a `broken spline', a set of separate cubic spline segments such that the first derivative is continuous everywhere but at the corners. The best locations of the corners will be found automatically during optimization of the spline points.

In our software a simple cubic spline fit is handled as a special case (one segment) of a broken spline. A segment through just two points degenerates automatically to a straight line.

Figure\,\ref{fig:extBSfit11} shows a broken spline fit (solid line) of a dose extinction curve. The small circles are input data. Squares are the spline points, and bold squares are the corners. What motivates the fit is that, once done, such descriptors of the curve as the 100\% level or the depth at the 50\% level can be found easily using standard routines. Working directly with the experimental data would be much messier, and subject to random noise.

Figure\,\ref{fig:BSfit11parms} shows the corresponding parameter file. Line 1 is the rms deviation of the fit, here in pC/MU. Line 2 gives the number of segments followed by the index of the last point in each segment. There follow arrays of spline point $x$, $y$ and $s$ values. $s$ is the second derivative, found during spline initialization, and 0 flags an endpoint or corner. Code to evaluate the fit at arbitrary $x$ using these arrays can be found in function WriteBSparms of module BSfit.FOR in folder \bs BGware\bs source at \BGsite\ .

\subsubsection{Managing Multiple Fits}
In a complex analysis it is necessary to preserve the parameters of many essentially unrelated fits and to use (for instance) a calibration based on data taken (let us say) 09MAR11 to interpret data taken 15JUN11. To handle this systematically we write the parameters of any fit of lasting value, be it polynomial or broken spline, to a text file in the subfolder of the appropriate date. Thus \bs diodeArray\bs 09MAR11\bs WEPLparms.txt contains the parameters of WEPL(t$_\mathrm{RMS}$) as determined on that date, and a similar file from another date could easily be substituted for comparison. The same machinery is used by a procedure to recall any fit performed earlier in that job, or several months ago. 
  
\subsection{Directory Structure Assumed by ArrayRMS}
The root directory is \bs diodeArray. Each run date is associated with a subdirectory e.g. \bs 15JUN11\,. On launch, the program prompts for the subdirectory because it needs that information to find the initialization file e.g. \bs diodeArray\bs 15JUN11\bs ArrayRMS.INI . If you are doing multiple runs you can create a shortcut to ArrayRMS.EXE and enter the path e.g. \bs diodeArray\bs 15JUN11 as the first parameter in the `target' line of the shortcut. That will allow you to answer the prompt with an empty return, rather than typing the path each time.

The program gets ArrayRMS.INI from the working subdirectory and puts all its output (mostly text files) there. Data files from the data logging system were, however, handled two quite different ways depending on when they were taken.

Through 28APR11 we saved data files by hand in the working subfolder, naming them according to their time stamp e.g. 071835.txt (which stands for 07:18:35, PM implied). Data files were distinguished from all othe TXT files by the fact that their name was a pure number.

After 28APR11 we used the automatic file-exporting machinery of ENCORE (the data logging program) and therefore, its directory structure, for raw data files. The path for e.g. 14JUN11 is
\begin{center}\bs Users\bs Bernie\bs Documents\bs Encore Data\bs 06142011\_024721\_PM\end{center}
and it contains two files, Voltage.txt (the 12-channel ADC) and Counter.txt (the beam monitor counter). Knowing the run date from the working path, the switches between the two schemes automatically. Eventually the earlier structure will fall away. For the present analysis we needed data of both kinds. 

\subsection{Using ArrayRMS.FOR}
The program reads an initialization file ArrayRMS.INI from the working directory. This is a self commented text file (Figure\,\ref{fig:iniFile}). Only the left-hand numbers or character strings have any effect; the rest is commentary. All entries are required as placeholders even if not all are used in a given run. The INI file is only used as far as the bottom line, so space below can be used for arbitrary comments, saving data and so on (Figure\,\ref{fig:iniFile}). 

The first entry is the task to be performed. At present there are four tasks, as follows.

\subsubsection{MAKELIST}
This task creates a file runList.txt having a header followed by one line for each data file, in order of time. The entries are 1) a serial run number, 2) the time stamp, 3) the elapsed minutes measured from the first run and 4) the DCEU counts obtained from the corresponding Counter.txt . This task is meant to create a convenient base to which information not `known' to the data logging system can be appended by hand.

For instance, the EXTINCT task to be described requires the beam range for each run. We ran MAKELIST, saved runList.txt as rangeList.txt (do this first!) and then added a `range' column. depthList.txt (for the 09MAR11 experiment) was created similarly.

There may be data which you wish provisionally to omit from analysis without, however, deleting them. It is only necessary to slightly spoil the directory name, changing, for instance, 06152011 to X6152011, to have MAKELIST omit those data. Such decisions should be made early as all subsequent tasks must apply to exactly the same set of runs.   

\subsubsection{PRELIM}\label{sec:prelim}
This is the longest task and provides generic pre-analysis for all subsequent tasks. For each run it reads the raw data and, for each channel, computes the DC baseline $v_\mathrm{corr}$ (Section\,\ref{sec:vcorr}), total charge, and the number of bursts. If the number of bursts is the number expected it then computes $\sigma_t$, skewness and kurtosis burst by burst, as well as the mean value, and the rms deviation from the mean, of each. After finishing all the runs it writes a number of text files, described below, each giving some statistic or measurement and its estimated error, by run (rows) and channel (columns). In most cases each row begins with the time stamp and the run's `minutes into experiment'. Thus standard graphics programs can import these files and plot any statistic for any channel as a function of time. Other tasks retrieve data from these text files for further analysis as will be described later. Figure\,\ref{fig:rmsWidth} shows a fragment of rmsWidth.txt as an example.\footnote{~Channel\,4 was dead.}

The text files produced by PRELIM are as follows:
\begin{description}
\item[MU.txt] lists the monitor units for each runs. (Counts per MU is a parameter in the INI file.)
\item[vcorr.txt] lists $v_\mathrm{corr}$, the average DC correction voltage (mV). The second set of twelve columns list, for each channel, an estimated rms width of the baseline voltage distribution. This is not computed directly, since it would be dominated by real signal in cases where the real signal could not be excluded (the number of bursts found was incorrect). Therefore the number given here is the $\sigma$ of a Gaussian that has the same width at the 90\% point (or whatever level is given in the INI file) as the observed spectrum has (see Figure\,\ref{fig:freq053_504_3_1_90}).
\item[nBursts.txt] lists, for each run and each channel, the number of bursts detected, or an error code if that was not the number expected. The burst finding algorithm could probably be improved. For instance, in the 15JUN11 experiment the correct number of bursts was found only 80\% of the time. That run had a lot of dose extinction measurements where, of course, the bursts get weaker as beam range is reduced. Burst finding does not, however, appear to be an issue in scout beam measurements where the beam has adequate range.
\item[pCpMU.txt] lists the total picoCoulombs (obtained by integrating the entire 2\,s trace) per monitor unit (delivered in the same 2\,s). The second set of twelve columns is supposed to give the rms deviation in the first, when a statistic is computed burst by burst. Since GLOBAL mode for total charge (recommended) does not work burst-by-burst, no error estimate is possible and those columns are zero.
\item[rmsWidth.txt] lists $\sigma_t$ (ms). Since that is burst-by-burst (by definition) an error estimate is possible and that is given in the second set of twelve columns.   
\item[skew.txt] lists the skewness and its error. 9.9999 is used as a placeholder if the skewness is out of bounds.
\item[kurtosis.txt] lists the kurtosis and its error. 99.999 is used as a placeholder if the kurtosis is out of bounds.
\end{description}
You can analyze any contiguous subset of runs using `min and max runs' in the INI file. The files just described will be overwritten each time. In the special case min = max (analyze one run only) PRELIM will write a (very long) file rawData.txt for that run (format is obvious) and stop, without overwriting the other text files. That gives you a chance to graph the raw data for any run to look for anomalies.

\subsubsection{TRMSCAL}
Scout beam depth-dose data for this task are obtained as described in Section\,\ref{sec:09MAR11procedure}. This task calibrates the scout beam, finding WEPL$_\mathrm{TRMS}$ = WEPL($\sigma_t$) and the `natural' functions skewness $s(\sigma_t)$ and kurtosis $k(\sigma_t)$. Depth-dose distributions are also produced as a reality check.

\subsubsection{EXTINCT}
This task compares scout beam data with dose extinction data obtained as described in Section\,\ref{sec:15JUN11procedure}. The files for figures and tables of Section\,\ref{sec:15JUN11} were produced with this task. \bs extBSfitIJ.txt is input for a graph of the broken spline (BS) fit to configuration I channel J (Figure\,\ref{fig:extBSfit11}). \bs extFitI.txt gives the BS fit points for configuration I, all channels (cf. Figure\,\ref{fig:extinction1}). \bs scoutResults.txt has data for Figure\,\ref{fig:pBoth}. \bs DEresults.tex lists data in \LaTeX\ format for Table\,\ref{tbl:DEresults}. \bs scoutResults.tex lists data in \LaTeX\ format for Table\,\ref{tbl:scoutResults}. Finally \bs dxxVweplI.txt has data for correlation plots (Fgures\,\ref{fig:dxxVwepl1} through \ref{fig:dxxVwepl3}). 

\section{09MAR11 Run and Analysis}\label{sec:09MAR11}

\subsection{Experimental Procedure}\label{sec:09MAR11procedure}
The goal of this run was to determine  $d(\sigma_t)$ = WEPL($\sigma_t$) = WEPL$_\mathrm{TRMS}$ = WEPL. With the same data it is also possible to determine the natural skewness and kurtosis $s(\sigma_t)$ and $k(\sigma_t)$.

The 16-diode array (DFLR1600 diodes, Figure\,\ref{fig:diodeArray16}) was flattened as well as possible, and mounted perpendicular to the beam axis in a CRS water tank. The CRS circuitry was used for positioning only. The diode array had to be shielded with external aluminum (household) foil. Only 8 channels of the amplifier were available so measurements were done first with diodes 1--8 and then with diodes 9--16. Because of various electronic problems only six diodes, 10--15, ultimately proved useful.

All files were taken with the scout beam. Our notes from this run are missing, but on 15JUN11 the scout beam parameters were: jaws 18$\times$18 / FS 3,5,9 / second scatterer 3 / modulator track 7 / BCM file a6mg / stop digit 220 / range at nozzle entrance 24.62 / current at cyclotron exit 5\,nA. The diode array was moved from 13 to 19\,cm depth (after correction for the CRS offset) in 0.2\,cm steps. One file comprising ten bursts (one second) was taken at each position, at 100\,Ksamples/s/channel. Note: the 0.01\,$\mu$F filter capacitor in the last amplifier stage (Figure\,\ref{fig:12chanAmplifier}) was not present for this run.

\subsection{Depth-Dose Distributions}
Figure\,\ref{fig:depthDose} shows depth-dose distributions from diodes 10--15. The dashed line at 18.3\,cm is the upper limit of the useful fitted region (next section), well beyond the knee as before \cite{BGwepl}. Interestingly, the distributions are flat, as with the PTW diode \cite{BGwepl}, without the peaking at low proton energies (large depths) that might be expected \cite{AMKdiode}. The factor 1.3$\times$ between the most and least sensitive diodes is unexceptional. The quality of these distributions, despite this run having no beam monitor, suggests the beam current is quite constant if nothing is changed.

\subsection{Function $d(\sigma_t)$ = WEPL($\sigma_t$) = WEPL$_\mathrm{TRMS}$}
Figure\,\ref{fig:WEPLfit} shows the fit to measured $d(\sigma_t)$ for diodes 10--15 using the procedure discussed earlier (Section\,\ref{sec:poly}). The global spread of data points about the fit is 0.079\,cm rms. Examination of the residuals (Figure\,\ref{fig:WEPLres}) reveals that part of the spread at each $d$ is non-random. Either the array was non-flat, or the effective depth of the diodes, due to the overburden of epoxy or possibly their orientation, introduces an effect of order 1\,mm. That suggests a more precise fit could be obtained with more attention to orthogonality to the beam, array flatness and diode orientation and, of course, more diodes. 

Since there is no limit to dose in these measurements, every means should eventually be taken to determine $d(\sigma_t)$ and the natural skewness and kurtosis with statistical and systematic errors that are negligible compared with those of in-phantom or in-patient measurements. In this report we assume they are comparable.

\subsection{Natural Skewness and Kurtosis}
Figure\,\ref{fig:natSkewFit} shows $s(\sigma_t)$ with a spline fit and Figure\,\ref{fig:natKurtFit} shows $k(\sigma_t)$. There is considerable scatter in both at low $\sigma_t$. However, the 15JUN11 analysis will use the region above 15\,ms which is quite flat and well fit.

\section{15JUN11 Run and Analysis}\label{sec:15JUN11}

\subsection{Preliminary Cable/Diode Array Tests}
The 12-channel amplifier has three DB9S input connectors each carrying four pairs and ground, instead of a single input connector. This was done to simplify internal wiring. However, the two 12-diode arrays were each furnished with a single DB25 ($12\times2+1$) connector. To accommodate the amplifier a $\hbox{DB25}\rightarrow 3\times\hbox{DB9}$ adapter cable (16\,inches to $3\times18$\,inches) was also provided. Unfortunately this increases total cable length, and rms noise is directly proportional to input cable capacitance. It also uses an extra pair of connectors, and connectors are a common source of problems in low level applications. However, it was too late to change back to simpler cabling. 

On 14JUN11, a day before the run, we bench tested various cable/diode combinations: open amplifier, adapter only, adapter plus diode array and so on. We used the IOtech\;6220 with a dummy trigger to collect data and we used ArrayRMS.FOR for analysis, here limited to finding the DC offset and rms noise (including 60\,Hz pickup) in each channel. We found that, even though the cables, connectors and diode arrays were nominally shielded, external aluminum foil shielding was needed to reduce 60\;Hz pickup to a level comparable to or below amplifier white noise. (In these tests the cables and diodes were lying on the bench, but we had removed any obvious 60\;Hz related sources such as transformers and fluorescent lamps. The 60\,Hz environment of a treatment room is different, of course, and probably worse due to multiple electronic devices.)

The results are summarized in Figure\,\ref{fig:vRMS14JUN11}. Large or off-scale values represent runs where at least one element had no external shielding. Final results for the `yellow' diode array, fully cabled and externally shielded, as well as the `blue' array, are at 30 and 60 minutes respectively. The bare amplifier is at $t=0$, and the points labeled `adapter' show the noise introduced by the adapter cable alone. The built in shielding over the diodes themselves was adequate for the `yellow' array but the `blue' array required aluminum foil over the diodes as well as the cable. Once shielded, the rms noise was $\approx6$\,mV of almost white noise, with a fair bit of channel to channel variation as is evident.

I believe the `blue' array was used for the 15JUN11 run. Only eight of the twelve diodes were useful. One faulty channel was due to the adapter cable, another was pinned (probably one side grounded), and two of the diodes gave much less signal than the good eight. 10--20\% interdiode variation in sensitivity, probably due to differences in active Si mass, seems to be normal. 

\subsection{Experimental Procedure}\label{sec:15JUN11procedure}
This run used a pelvic phantom. The phantom was aligned in the beam, the externally shielded diode array was inserted around the rectal balloon and the balloon was filled. X-rays were taken to locate the diodes. However, that aspect of the data (discussed in ST) is not covered in this report, which concerns itself only with what we can learn from diode signals regardless of where each diode is actually located. 

Three configurations were studied in the following order: 1) open field (OF), beam on phantom without aperture or range compensator; 2) beam on phantom through aperture and range compensator (AP\,+\,RC) and 3) a solid water phantom (SWP). For SWP, the diode array was flattened as well as possible.

Each run comprised 20 bursts (2\,s) and was repeated, except for scout beam runs which were done four times, for a total of 163 voltage files. (One repetition failed because of a room search timeout that was not noticed at the time.) OF was set up and runs 1--4 taken with the scout beam. Then DE runs 5--72 were taken with the beam range reduced from 15.5 to 12.2\,cm in 1\,mm steps. AP\,+\,RC was set up and DE runs 73--119 were taken with the range increased from 13.0 to 16.0\,cm in varying steps, mostly 1\,mm. Three sets of scout runs were than taken: 120--123 with AP\,+\,RC, 124--127 repeating OF, and 128--131 with SWP. Finally DE runs 132--163 were made with SWP decreasing the beam range from 15.5 to 13.5\,cm mostly in 2\,mm steps. The whole experiment took about three hours.

Figure\,\ref{fig:MU} serves as a sort of run outline, and shows monitor units (MU) delivered for each run.  

\subsection{Dose Extinction}
Unlike depth-dose (DD) measurements, dose extinction (DE) measurements are uncommon and their interpretation is not completely obvious. The essential problem can be put this way: Suppose we take DD data on an SOBP (flat by design) as usual, scanning a dosimeter in depth $z$ along the beam axis. We then fix the dosimeter at some $z$ and take DE data, exposing the water tank to successive range modulated beams of decreasing ranges. It seems obvious the DD and DE data must be closely related. Can DE results be predicted exactly given the DD results? For instance, is $d_{80}$, the depth at the 80\% point, the same in the two methods? For that matter, how should the 100\% dose level be defined?

We have not yet formally thought through the entire problem. The dependence of beam monitor output factor on beam parameters should come in, because DE involves different beams whereas DD does not. (ST puts in the output factor and claims its effect is negligible, but the overall reasoning is not explained.) A fluence factor should also come in, because detector $z$ is fixed in DE but not in DD.

In this note we will ignore these issues and analyze DE curves as though they were DD measurements. The 100\% dose level, for determining $d_{80}$ and the like, will simply be defined as the mean value in the flat region (Figure\,\ref{fig:extBSfit11}).

Figures\,\ref{fig:extinction1}--\ref{fig:extinction3} are DE curves for OF, AP+RC and SWP respectively with associated broken spline (BS) fits. The figures are purposely not normalized, and they are plotted to the same scale to emphasize differences. Scout beam points are included; they are valuable in defining the 100\% level. As expected, the dose rises sooner for OF than for AP+RC: there is less material in the beam. Dose onset is deeper and more tightly clustered for AP+RC: the range compensator is doing its job. The curves are very nearly identical for SWP except for $\approx30\%$ spread in the plateau, undoubtedly due to diode sensitivity. And the curves fall very nearly to zero, indicating the DC baseline computation is correct.

Table\,\ref{tbl:DEresults} summarizes DE results. Columns 2--6 are details of the BS fit: number of segments required, points in each segment and goodness of fit (rms deviation of measured points from the fit, $\approx2\%$ of $D_{100}$). Column 7 lists $D_{100}$. The next four columns list $d_\mathrm{xx}$ at several levels (whenever possible). The final column lists $(d_{80}-d_{20})$, the width of the rise.
For AP+RC, channels 4 and 6 have anomalously large $(d_{80}-d_{20})$ (see also Figure\,\ref{fig:extinction2}). That will become significant in the next section.

\subsection{Scout Beam}
Table\,\ref{tbl:scoutResults} gives results for the scout beam measurements. Column 1 lists the channel (channels 1--9 listed as 1--8, omitting defective channel 4). Columns 2 lists $\sigma_t$ and column 3 lists WEPL from $d(\sigma_t)$. Columns 4 and 5 list excess skewness (\%) and the estimated error in mean excess skewness (\%) computed per Section\,\ref{sec:p} . Columns 6 and 7 treat kurtosis likewise. Columns 8--10 list the corresponding values of $p_\mathrm{ES}$, $p_\mathrm{EK}$ and their product $p_\mathrm{both}$ (Section\,\ref{sec:p}). The $p$ values are strikingly lower for the two channels in AP+RC, 4 and 6, that show independent evidence of range mixing.  

\subsection{Correlation Between $d_{80}$ and WEPL($\sigma_t$)}
Figure\,\ref{fig:dxxVwepl1} shows the correlation between $d_{80}$ and WEPL($\sigma_t$) for OF. The correlation is excellent for all channels except 4 and 6 where $d_{80}$ could not be found (see Figure\,\ref{fig:extinction1}). It is markedly worse for $d_{90}$ (open circles). Discrepancies, of order 0.1\,cm, could be due to any of a number of causes. Figure\,\ref{fig:dxxVwepl2} for AP+RC (note change of scale) shows equally good results except, again, for channels 4 and 6 which reveal range mixing in their DE distribution (Figure\,\ref{fig:extinction2}). Even for SWP, where the diodes are nominally at the same depth, there is evidence of correlation (Figure\,\ref{fig:dxxVwepl2}, scale changed again). Evidently some of the spread corresponds to real differences in effective depth, as we have already seen with a similar array in a previous run (Figure\,\ref{fig:WEPLres}).   

\subsection{Range Mixing and $p$}
Figure\,\ref{fig:pBoth} plots $p_\mathrm{both}$ from Table\,\ref{tbl:scoutResults} for all three configurations. AP+RC and SWP present no problems: $p_\mathrm{both}$ is in the 1--100\% range except for channels 4 and 6 of AP+RC, which show clear evidence of range mixing (Figure\,\ref{fig:extinction2}).

OP is less clear-cut (see also Figure\,\ref{fig:extinction1}). Channels 1, 5 and 7 show sharp DE curves and correspondingly large $p$. Channels 2, 4 and 6 have ambiguous DE curves and small, though not very small, $p$. Channel 8, with its relatively sharp DE curve yet low $p$ is puzzling. Finally channel 3 with its very low $p$ could be consistent; its DE curve was not fully mapped, but looks like it could have a non-sharp falloff.

To emphasize just how subtle are the effects we have been exploring statistically, Figure\,\ref{fig:raw123ch2ch4} compares five bursts each from scout beam file \#123, AP+RC, channel 2 (no mixing) and channel 4 (mixing). It helps a great deal, of course, that we have 80 of each to work with.

\section{Discussion}\label{sec:Discussion}

We have learned a lot since the first experiment with the PTW. Ordinary diodes work well, and have the same $d(\sigma_t)$ as the PTW. In either our results or ST, SB and DE measurements are well correlated. The fact that this holds up in the phantom suggests that orientation is not a major issue. There is strong indication that range mixing can be detected by looking at skewness and kurtosis, but that clearly needs further work.

\subsection{Hardware Improvements}
The present method of fabricating diode arrays needs improvement. The yield of good channels is too low, shielding is inadequate for some reason, and it is difficult or impossible to repair a bad channel. We should explore packaging each diode separately, shielded and waterproofed, connected to its own shielded twisted pair. They could then be bench tested and wired in quads to the DB9 connectors. The rectal balloon would be inflated and the diodes affixed with a thin transparent film such as 3M Tegaderm$^\mathrm{TM}$. We feel this approach should at least be attempted.

About 3\,feet (1\,m) of cable is required to get the amplifier out of the beam. Noise is proportional to cable capacitance. The precision of $\sigma_t$, $s$ and $k$ is noise limited and less noise means potentially less dose. We have measured the following pF/foot: Hassan adapter cable, 48 ; Belden\,9501 (ordinary shielded twisted pair), 43 ; New England Wire Technologies (NE) N13--44B--4658--1, 27 ; NE N13--50B--4608--0, 19 . The last named is not only best for capacitance but also the thinnest and most flexible; it would be easy to get twelve cables out of the rectum as long as they were not tightly bundled. It costs \$1632/1000$'$ (minimum order, 10/10/2011 quotation) or \$1.63\,/ft, not particularly expensive.

Whoever builds the diode array needs more than a simple digital volt/ohm meter to test the diode and the integrity of the shielding. One should have at least a single channel current-to-voltage amplifier, a digital scope, and a $\approx10\,\mu$\,Ci radioactive source. Timely bench tests of the full setup will avoid wasted time during the run and suboptimal data. 

Ultimately it may be possible to locate the IOtech\,6220 in the treatment room, using only short coaxial cables and letting the Ethernet connection be the long run to the TCR. This will undoubtedly work, but leaves the 6220 exposed to neutrons which (if experience is any guide) will cause occasional upsets which may or may not be cured by re-booting. In principle the 6220 can be rebooted remotely using Encore, but it is not obvious that will work for an arbitrary upset due to radiation. We wished to avoid these potential problems, as well as possible permanent damage to the 6220, during these preliminary tests. In the long run it should be explored as it would simplify the setup considerably.

\subsection{Further Experiments}
Having validated the SB method by using DE, we don't need further DE experiments just now. The most pressing need is to explore range mixing in a controlled way. We should first calibrate $d(\sigma_t)$ and the natural functions $s(\sigma_t)$ and $k(\sigma_t)$ using our standard scout beam with a transverse diode array in a water tank. A half-beam `bone' slab (perhaps with a tapered edge) could then be introduced upstream to introduce range mixing with a known off-axis dependence. Preliminary Monte Carlo or other theoretical dosimetric studies should be done to ensure that we can produce range mixing of both kinds \cite{Urie1986}: continuous (where the distal edge is merely broadened) and bimodal (where two distal edges are evident). The experiment would use the scout beam alone, and the accuracy of the theoretical prediction could be tested using the measured depth-dose. With good planning this should take no more than a shift.

After a range mixing experiment, another SB and DE measurement in the phantom would be useful as a final test of the technique before attempting a patient. From here on AP+RC is probably sufficient. The proposed treatment procedure is actually a two-point DE measurement. The SB measurement calibrates the diode sensitivity and provides a range correction, while the subsequent treatment (during which we will still take data) measures the dose at the balloon when the correct treatment range is used. If all is understood, all should hang together, and this should be verified with a mock treatment of a phantom. Somehow, of course, we would have to introduce an `unkown' range error to be compensated; perhaps a small addition to the RC whose water equivalent thickness is unknown to the experimenters.

\subsection{Grand Summary} 
We continue to make progress in understanding the technique. No show-stoppers yet.

\subsection{Acknowledgements}
The work described in this note is largely mine, but owes everything to work and ideas from each collaborator: Hassan Bentefour, Ethan Cascio, Hsiao-Ming Lu, Damien Prieels and Shikui Tang. The hardware part of this work was performed under contract with IBA. I thank Harvard University for its support through the Physics Department and the Laboratory for Particle Physics and Cosmology.

\clearpage

\begin{thebibliography}{10}

\bibitem{hmLuInVivo}
Hsiao-Ming Lu, `A potential method for {\it in vivo} range verification in
  proton therapy treatment,' Phys. Med. Biol. {\bf53} (2008) 1413-1424.

\bibitem{BGwepl}
B. Gottschalk, S. Tang, E.H. Bentefour, E.W. Cascio, D. Prieels and H-M Lu,
  `Water equivalent path length measurement in proton radiotherapy using time
  resolved diode dosimetry,' Med. Phys. {\bf38(4)} (2011) 2282-2288.

\bibitem{ST09MAR11}
Shikui Tang, `Data analysis of experiment on March 9 and 10, 2011,' technical
  memo (March 2011).

\bibitem{ST15JUN11}
Shikui Tang, `Data analysis of experiment on June 15, 2011,' technical memo
  (July 2011).

\bibitem{STrangeMixingPhantom}
Shikui Tang, `The trms-WEPL relation in pelvic phantom: effects of range
  mixing,' unpublished technical note (2011).

\bibitem{STrangeMixing}
Shikui Tang, `Effects of range mixing on t$_\mathrm{rms}$-WEPL relationship,'
  unpublished technical note (2011).

\bibitem{Urie1986}
Marcia Urie, Michael Goitein, W.R. Holley and George T.Y. Chen, `Degradation of
  the Bragg peak due to inhomogeneities,' Phys. Med. Biol. {\bf31(1)} (1986)
  1-15.

\bibitem{nr}
W.H. Press, B.P. Flannery, S.A. Teukolsky and W.T. Vetterling, ``Numerical
  Recipes: the Art of Scientific Computing,'' Cambridge University Press
  (1986).

\bibitem{bevington}
Philip R. Bevington, ``Data Reduction and Error Analysis for the Physical
  Sciences,'' McGraw-Hill (1969).

\bibitem{AMKdiode}
A.M. Koehler, `Dosimetry of proton beams using small silicon diodes,' Rad. Res.
  Suppl. {\bf7} (1967) 53-63. In unpublished Errata the author presents a
  corrected Figure 4b consistent with an output change of $+0.32\%/\;$\degr C.

\end{thebibliography}

\listoftables

\clearpage
\listoffigures

\begin{table}[p]
\begin{center}\begin{tabular}{rrrrrrrrrrrr}
\multicolumn{12}{l}{Open Field (OF):}\\~\\
\multicolumn{1}{c}{chan}&           
\multicolumn{1}{c}{segs}&           
\multicolumn{1}{c}{$n_1$}&           
\multicolumn{1}{c}{$n_2$}&           
\multicolumn{1}{c}{$n_3$}&           
\multicolumn{1}{c}{$\sigma_\mathrm{fit}$}&           
\multicolumn{1}{c}{$D_{100}$}&           
\multicolumn{1}{c}{$d_{90}$}&           
\multicolumn{1}{c}{$d_{80}$}&           
\multicolumn{1}{c}{$d_{50}$}&           
\multicolumn{1}{c}{$d_{20}$}&
\multicolumn{1}{c}{$d_{8020}$}\\           
\multicolumn{1}{c}{}&           
\multicolumn{1}{c}{}&           
\multicolumn{1}{c}{}&           
\multicolumn{1}{c}{}&           
\multicolumn{1}{c}{}&           
\multicolumn{1}{c}{pC/MU}&           
\multicolumn{1}{c}{pC/MU}&           
\multicolumn{1}{c}{cm}&           
\multicolumn{1}{c}{cm}&           
\multicolumn{1}{c}{cm}&           
\multicolumn{1}{c}{cm}&           
\multicolumn{1}{c}{cm}\\           
  1&  3&  3&  4&  2&   9.08&  385.0&  14.309&  14.219&  13.929&  13.581&   0.638\\
  2&  3&  3&  4&  2&   5.21&  314.2&  13.707&  13.440&  13.020&  12.643&   0.797\\
  3&  2&  2&  4&  0&   4.77&  347.9&  13.372&  12.464&        &        &        \\
  4&  1&  3&  0&  0&   3.68&  344.9&  13.562&        &        &        &        \\
  5&  3&  3&  4&  2&   6.19&  335.1&  14.091&  13.994&  13.697&  13.391&   0.603\\
  6&  1&  3&  0&  0&   4.02&  295.0&  13.491&        &        &        &        \\
  7&  3&  3&  4&  2&   6.49&  355.6&  13.955&  13.797&  13.504&  13.226&   0.571\\
  8&  2&  2&  4&  0&   5.26&  353.3&  13.639&  12.866&  12.561&  12.257&   0.609\\
\\~\\\multicolumn{12}{l}{Aperture + Range Compensator (AP\,+\,RC):}\\~\\
\multicolumn{1}{c}{chan}&           
\multicolumn{1}{c}{segs}&           
\multicolumn{1}{c}{$n_1$}&           
\multicolumn{1}{c}{$n_2$}&           
\multicolumn{1}{c}{$n_3$}&           
\multicolumn{1}{c}{$\sigma_\mathrm{fit}$}&           
\multicolumn{1}{c}{$D_{100}$}&           
\multicolumn{1}{c}{$d_{90}$}&           
\multicolumn{1}{c}{$d_{80}$}&           
\multicolumn{1}{c}{$d_{50}$}&           
\multicolumn{1}{c}{$d_{20}$}&
\multicolumn{1}{c}{$d_{8020}$}\\           
\multicolumn{1}{c}{}&           
\multicolumn{1}{c}{}&           
\multicolumn{1}{c}{}&           
\multicolumn{1}{c}{}&           
\multicolumn{1}{c}{}&           
\multicolumn{1}{c}{pC/MU}&           
\multicolumn{1}{c}{pC/MU}&           
\multicolumn{1}{c}{cm}&           
\multicolumn{1}{c}{cm}&           
\multicolumn{1}{c}{cm}&           
\multicolumn{1}{c}{cm}&           
\multicolumn{1}{c}{cm}\\           
  1&  3&  3&  4&  2&   6.27&  376.4&  15.135&  15.001&  14.713&  14.442&   0.559\\
  2&  3&  2&  4&  2&   6.00&  304.7&  15.381&  15.230&  14.960&  14.679&   0.551\\
  3&  3&  3&  4&  2&   5.71&  322.8&  15.274&  15.144&  14.821&  14.472&   0.673\\
  4&  3&  2&  4&  2&   4.45&  306.1&  14.898&  14.654&  14.011&  13.449&   1.205\\
  5&  3&  3&  4&  2&   6.01&  321.2&  14.542&  14.407&  14.140&  13.872&   0.535\\
  6&  3&  2&  4&  2&   3.91&  255.7&  15.566&  15.389&  14.851&  14.231&   1.159\\
  7&  3&  2&  4&  2&   7.71&  339.9&  14.309&  14.191&  13.907&  13.584&   0.607\\
  8&  3&  3&  4&  2&   6.73&  326.8&  14.335&  14.233&  13.962&  13.651&   0.582\\
\\~\\\multicolumn{12}{l}{Solid Water Phantom (SWP):}\\~\\
\multicolumn{1}{c}{chan}&           
\multicolumn{1}{c}{segs}&           
\multicolumn{1}{c}{$n_1$}&           
\multicolumn{1}{c}{$n_2$}&           
\multicolumn{1}{c}{$n_3$}&           
\multicolumn{1}{c}{$\sigma_\mathrm{fit}$}&           
\multicolumn{1}{c}{$D_{100}$}&           
\multicolumn{1}{c}{$d_{90}$}&           
\multicolumn{1}{c}{$d_{80}$}&           
\multicolumn{1}{c}{$d_{50}$}&           
\multicolumn{1}{c}{$d_{20}$}&
\multicolumn{1}{c}{$d_{8020}$}\\           
\multicolumn{1}{c}{}&           
\multicolumn{1}{c}{}&           
\multicolumn{1}{c}{}&           
\multicolumn{1}{c}{}&           
\multicolumn{1}{c}{}&           
\multicolumn{1}{c}{pC/MU}&           
\multicolumn{1}{c}{pC/MU}&           
\multicolumn{1}{c}{cm}&           
\multicolumn{1}{c}{cm}&           
\multicolumn{1}{c}{cm}&           
\multicolumn{1}{c}{cm}&           
\multicolumn{1}{c}{cm}\\           
  1&  3&  3&  4&  2&   9.03&  375.7&  14.997&  14.857&  14.579&  14.305&   0.553\\
  2&  3&  3&  4&  2&   6.46&  302.0&  14.970&  14.820&  14.534&  14.278&   0.542\\
  3&  3&  3&  4&  2&   7.71&  333.8&  14.994&  14.869&  14.590&  14.314&   0.555\\
  4&  3&  3&  4&  2&   7.10&  311.6&  15.001&  14.883&  14.597&  14.310&   0.574\\
  5&  3&  3&  4&  2&   7.05&  320.7&  14.935&  14.799&  14.540&  14.271&   0.529\\
  6&  3&  3&  4&  2&   6.99&  282.9&  14.935&  14.788&  14.522&  14.280&   0.508\\
  7&  3&  3&  4&  2&   6.87&  343.1&  15.025&  14.873&  14.586&  14.315&   0.558\\
  8&  3&  3&  4&  2&   8.56&  330.8&  14.923&  14.764&  14.489&  14.234&   0.530\\
\end{tabular}\end{center}
\caption{15JUN11 results from dose extinction.\label{tbl:DEresults}}
\end{table}
\clearpage

\begin{table}[p]
\begin{center}\begin{tabular}{rrrrrrrrrr}
\multicolumn{10}{l}{Open Field (OF):}\\~\\
\multicolumn{1}{c}{chan}&           
\multicolumn{1}{c}{$\sigma_t$}&           
\multicolumn{1}{c}{WEPL}&           
\multicolumn{1}{c}{ExcSkew}&           
\multicolumn{1}{c}{$\sigma_\mathrm{mean}$}&           
\multicolumn{1}{c}{ExcKurt}&           
\multicolumn{1}{c}{$\sigma_\mathrm{mean}$}&           
\multicolumn{1}{c}{$p_\mathrm{ES}$}&           
\multicolumn{1}{c}{$p_\mathrm{EK}$}&           
\multicolumn{1}{c}{$p_\mathrm{both}$}\\           
\multicolumn{1}{c}{}&           
\multicolumn{1}{c}{ms}&           
\multicolumn{1}{c}{cm}&           
\multicolumn{1}{c}{\%}&           
\multicolumn{1}{c}{\%}&           
\multicolumn{1}{c}{\%}&           
\multicolumn{1}{c}{\%}&           
\multicolumn{1}{c}{\%}&           
\multicolumn{1}{c}{\%}&           
\multicolumn{1}{c}{\%}\\           
  1&  17.19&  14.25&  0.961&  1.428& -0.379&  1.166&  50.11542&  74.49651&  37.33424\\
  2&  18.25&  13.47&  2.317&  1.082&  0.075&  0.676&   3.22427&  91.20343&   2.94064\\
  3&  19.40&  12.52&  3.258&  0.910&  0.856&  0.605&   0.03411&  15.71976&   0.00536\\
  4&  21.67&  10.30&  6.212&  0.776&  2.356&  0.577&   0.00000&   0.00453&   0.00000\\
  5&  17.49&  14.04& -0.023&  0.906& -0.182&  0.747&  97.94058&  80.71388&  79.05164\\
  6&  20.99&  11.01&  5.904&  0.925&  1.912&  0.704&   0.00000&   0.65700&   0.00000\\
  7&  17.72&  13.88&  0.898&  0.927& -0.206&  0.727&  33.27993&  77.65801&  25.84453\\
  8&  18.80&  13.02&  2.050&  0.789&  0.786&  0.629&   0.93286&  21.08239&   0.19667\\
\\~\\\multicolumn{10}{l}{Aperture + Range Compensator (AP\,+\,RC):}\\~\\
\multicolumn{1}{c}{chan}&           
\multicolumn{1}{c}{$\sigma_t$}&           
\multicolumn{1}{c}{WEPL}&           
\multicolumn{1}{c}{ExcSkew}&           
\multicolumn{1}{c}{$\sigma_\mathrm{mean}$}&           
\multicolumn{1}{c}{ExcKurt}&           
\multicolumn{1}{c}{$\sigma_\mathrm{mean}$}&           
\multicolumn{1}{c}{$p_\mathrm{ES}$}&           
\multicolumn{1}{c}{$p_\mathrm{EK}$}&           
\multicolumn{1}{c}{$p_\mathrm{both}$}\\           
\multicolumn{1}{c}{}&           
\multicolumn{1}{c}{ms}&           
\multicolumn{1}{c}{cm}&           
\multicolumn{1}{c}{\%}&           
\multicolumn{1}{c}{\%}&           
\multicolumn{1}{c}{\%}&           
\multicolumn{1}{c}{\%}&           
\multicolumn{1}{c}{\%}&           
\multicolumn{1}{c}{\%}&           
\multicolumn{1}{c}{\%}\\           
  1&  15.85&  15.11&  0.818&  1.527& -0.808&  1.103&  59.23814&  46.35600&  27.46044\\
  2&  15.55&  15.30& -0.438&  1.178&  0.047&  0.893&  71.01514&  95.81335&  68.04199\\
  3&  15.72&  15.19&  1.467&  0.924& -0.073&  0.696&  11.24245&  91.60463&  10.29860\\
  4&  16.95&  14.42&  2.414&  0.961&  2.266&  0.534&   1.19686&   0.00224&   0.00003\\
  5&  16.84&  14.49&  0.559&  1.058&  0.105&  0.621&  59.73556&  86.53895&  51.69452\\
  6&  15.77&  15.17&  2.439&  1.348&  2.365&  1.024&   7.03359&   2.09025&   0.14702\\
  7&  17.25&  14.21&  0.548&  0.798& -0.686&  0.723&  49.25312&  34.28473&  16.88630\\
  8&  17.15&  14.28& -0.664&  0.846&  0.180&  0.668&  43.26891&  78.75501&  34.07644\\
\\~\\\multicolumn{10}{l}{Solid Water Phantom:}\\~\\
\multicolumn{1}{c}{chan}&           
\multicolumn{1}{c}{$\sigma_t$}&           
\multicolumn{1}{c}{WEPL}&           
\multicolumn{1}{c}{ExcSkew}&           
\multicolumn{1}{c}{$\sigma_\mathrm{mean}$}&           
\multicolumn{1}{c}{ExcKurt}&           
\multicolumn{1}{c}{$\sigma_\mathrm{mean}$}&           
\multicolumn{1}{c}{$p_\mathrm{ES}$}&           
\multicolumn{1}{c}{$p_\mathrm{EK}$}&           
\multicolumn{1}{c}{$p_\mathrm{both}$}\\           
\multicolumn{1}{c}{}&           
\multicolumn{1}{c}{ms}&           
\multicolumn{1}{c}{cm}&           
\multicolumn{1}{c}{\%}&           
\multicolumn{1}{c}{\%}&           
\multicolumn{1}{c}{\%}&           
\multicolumn{1}{c}{\%}&           
\multicolumn{1}{c}{\%}&           
\multicolumn{1}{c}{\%}&           
\multicolumn{1}{c}{\%}\\           
  1&  16.26&  14.86& -1.251&  0.664& -0.730&  0.597&   5.96225&  22.11978&   1.31884\\
  2&  16.26&  14.87& -0.897&  0.953& -0.908&  0.597&  34.68203&  12.83770&   4.45237\\
  3&  16.13&  14.95&  0.443&  0.953& -0.497&  0.759&  64.24179&  51.25954&  32.93004\\
  4&  16.12&  14.96&  0.780&  0.991& -1.077&  0.743&  43.13163&  14.73860&   6.35700\\
  5&  16.25&  14.87& -0.265&  0.785& -0.485&  0.605&  73.58296&  42.26549&  31.10020\\
  6&  16.26&  14.87&  0.766&  0.800& -0.783&  0.771&  33.83253&  30.94997&  10.47116\\
  7&  16.19&  14.91&  0.163&  1.022& -0.328&  0.779&  87.29426&  67.35313&  58.79542\\
  8&  16.37&  14.80& -0.398&  0.906& -1.047&  0.656&  66.05752&  11.06139&   7.30688\\
\end{tabular}\end{center}
\caption{15JUN11 results from scout beam.\label{tbl:scoutResults}}
\end{table}
\clearpage

\clearpage

\clearpage

\begin{figure}[p]
\begin{center}\begin{verbatim}
\diodeArray\09MAR11\WEPLparms.txt:

 5
   13.09000      0.1677852    
   16.48685      -2.423201      -1.037287     -0.2816583      6.8953469E-02
  1.2367997E-02  2.9060086E-02  7.6808281E-02  4.2779777E-02  8.4668517E-02

code to read fit parameters:

READ (lun,*) np
READ (lun,*) xmid,slope
READ (lun,*) (pp(i),i=1,np)
READ (lun,*) (dp(i),i=1,np)

code to evaluate fitted WEPL at x:

WEPLfunc = Poly(slope*(x-xmid),np,pp)

equivalent formulas:
\end{verbatim}
\[\mathrm{WEPL}\;=\;\sum_{i=1}^\mathtt{np}\mathtt{pp(i)}\;u^{i-1}\]
\[u\;=\;\mathtt{slope}\times(\sigma_t-\mathtt{xmid})\]
\end{center}
\caption{Parameters of 5-term polynomial fit to WEPL($\sigma_t$), code to read them, code to evaluate fitted WEPL, and equivalent mathematical formulas. $\mathtt{dp(i)}$ are estimated errors in the polynomial coefficients.\label{fig:WEPLparms}}
\end{figure}
\clearpage

\begin{figure}[p]
\centering\includegraphics[width=4.44in,height=3.5in]{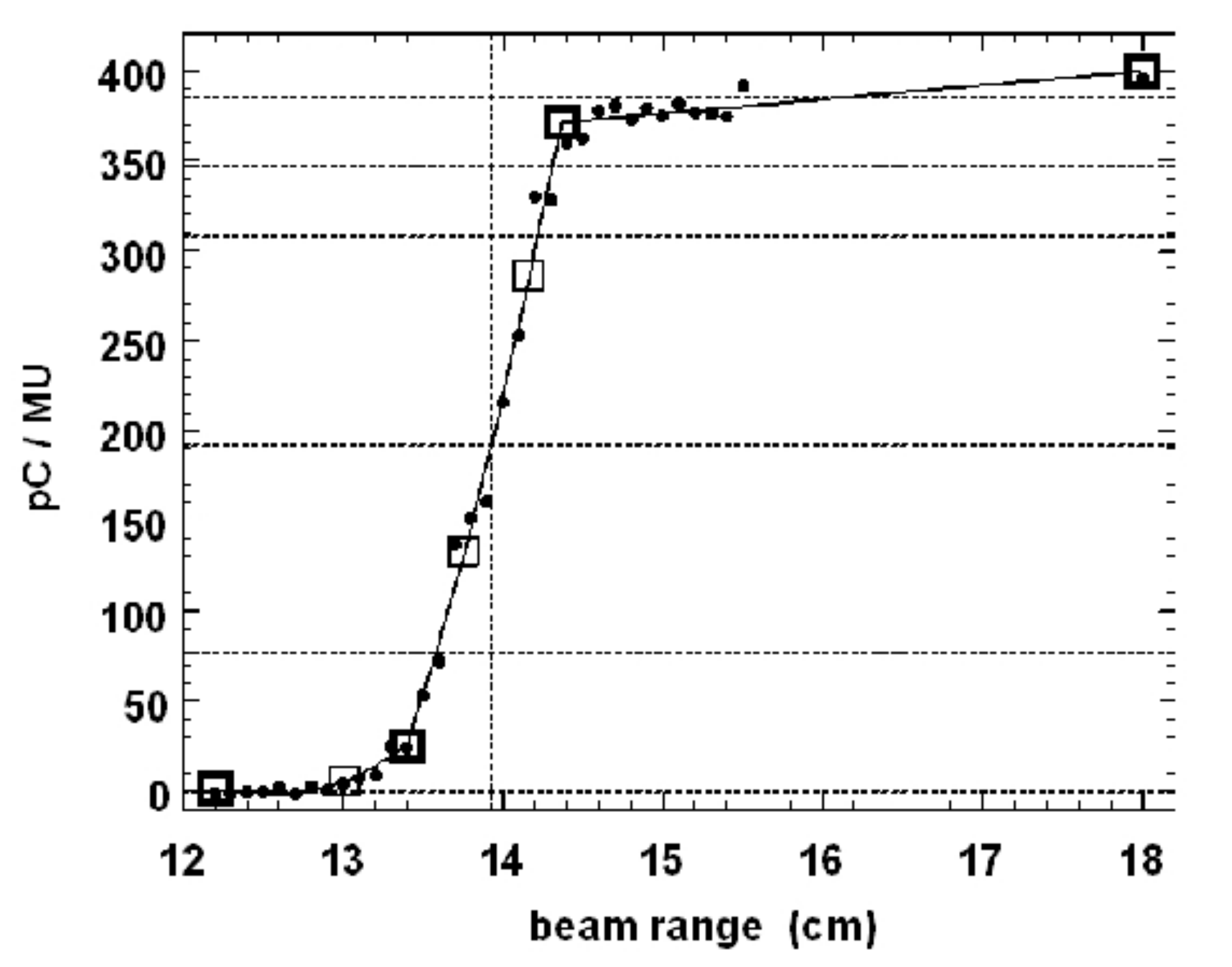} 
\caption{Three-segment broken spline fit to DE measurements (full circles) for OF, channel\,1. Open squares: spline points (adjustable parameters of the fit). Bold squares: corners, at which the slope is discontinuous and the second derivative is zero. The 100, 90, 80, 50, 20 and 0\% levels are shown, as is $d_{50}$.\label{fig:extBSfit11}}
\end{figure}
\begin{figure}[p]
\begin{center}\begin{verbatim}
\diodeArray\15JUN11\BSfit11.txt:

   9.078171    
  3     3   6   7
   12.20000       13.00570       13.39971       13.75833       14.16585    
   14.36582       18.00000    
  0.7568561       5.222276       24.61742       132.2755       285.8431    
   370.9038       399.1067    
  0.0000000E+00   109.2357      0.0000000E+00   259.4211       152.7311    
  0.0000000E+00  0.0000000E+00
\end{verbatim}\end{center}
\caption{Parameters file for broken spline fit to DE measurements for OF, channel\,1.\label{fig:BSfit11parms}}
\end{figure}
\clearpage

\begin{figure}[p]
\begin{center}\begin{verbatim}
PRELIM        preliminary analysis, runs in range
GLOBAL        charge/burst: 'GLOBAL' sum or 'byBurst'
VMEAN         use VMEAN or VMODE for voltage baseline
.FALSE.       show frequency distributions
1  163        min,max runs to analyze
111011111000  good channels
1             selected channel for printout
300           counts/monitor unit
0.5           amplifier volts/nA
0.504         channel width 2nd pass (mV)
3  1          smoothing rms bins, # passes
0.90          fraction of max freq for vMean calculation
.995          fraction of mod period for burst search
100           mod period (ms)
7             vrms multiplier to qualify as pulse
18.3          depth cut (cm)
-4.0          depth correction
5             # terms in d(sigma) polynomial           
20            # bursts expected
3             # extinction sets
1   4         run# range for diode normalization  OF
1  72         run# range for extinction
120  123      run# range for diode normalization  AP+RC
73   123      run# range for extinction
128  131      run# range for diode normalization  SWP
128  163      run# range for extinction
------------------------------------------------------------
tasks:
MAKELIST      make run list with DCEU counts
PRELIM        preliminary analysis, runs in range
RMSCAL        calibrate depth v. rms burst width
EXTINCT       plot dose extinction
\end{verbatim}\end{center}
\caption{Initialization file \bs diodeArray\bs 15JUN11\bs ArrayRMS.txt.\label{fig:iniFile}}
\end{figure}

\begin{figure}[h]
\begin{center}\begin{verbatim}
 file  relMin  rmsCH1  rmsCH2  rmsCH3  rmsCH4   ....   drmCH1  drmCH2  drmCH3  drmCH4
084730   0.000   17.13   18.34   19.47    0.00  ....      0.33    0.34    0.28    0.00
084808   0.633   17.30   18.24   19.48    0.00  ....      0.44    0.30    0.34    0.00
084820   0.833   17.18   18.19   19.47    0.00  ....      0.47    0.33    0.29    0.00
084831   1.017   17.17   18.26   19.23    0.00  ....      0.48    0.37    0.21    0.00
\end{verbatim}\end{center}
\caption{Fragment of \bs diodeArray\bs 15JUN11\bs rmsWidth.txt.\label{fig:rmsWidth}}
\end{figure}
\clearpage

\begin{figure}[p]
\centering\includegraphics[width=3.5in,height=3.5in]{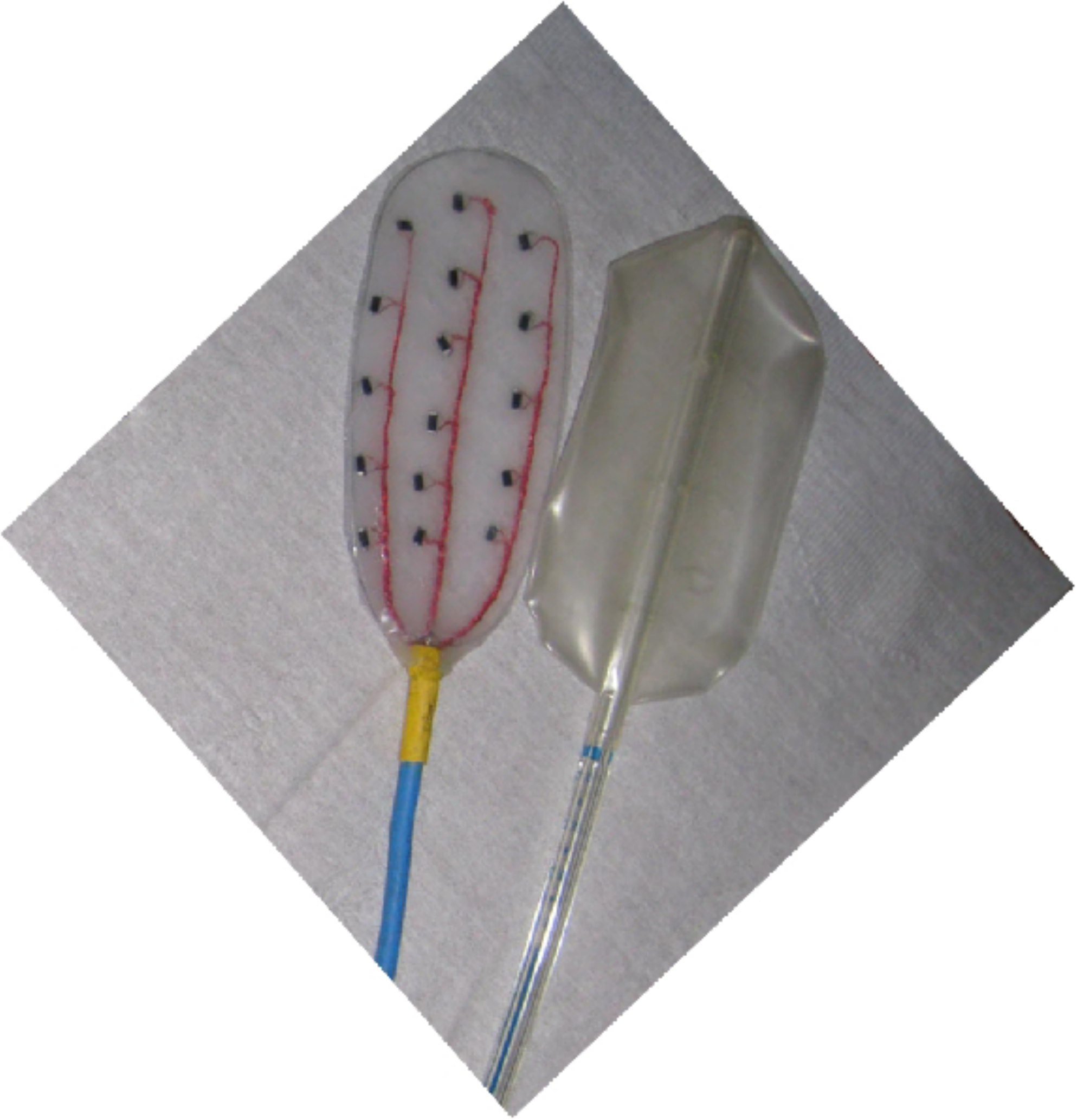} 
\caption{16-diode array (aluminum foil shield not shown) and rectal balloon. (Photograph by courtesy of H.\,Bentefour, IBA.)\label{fig:diodeArray16}}
\end{figure}
\begin{figure}[p]
\centering\includegraphics[width=4.55in,height=3.5in]{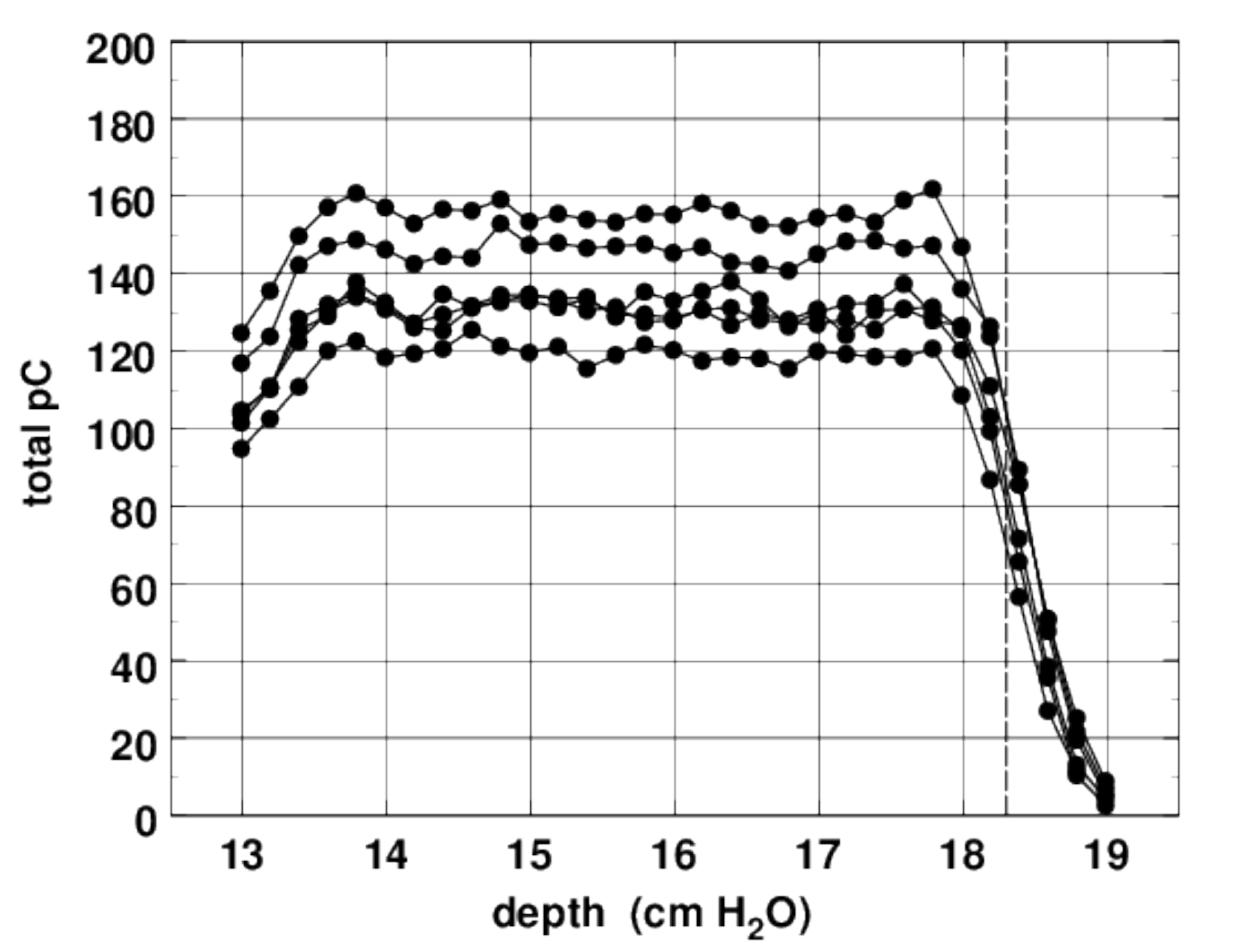} 
\caption{Depth-dose distributions for the six diodes used in the WEPL calibration. Dashed line: upper limit (18.3\,cm) of fitted region.\label{fig:depthDose}}
\end{figure}
\begin{figure}[p]
\centering\includegraphics[width=4.55in,height=3.5in]{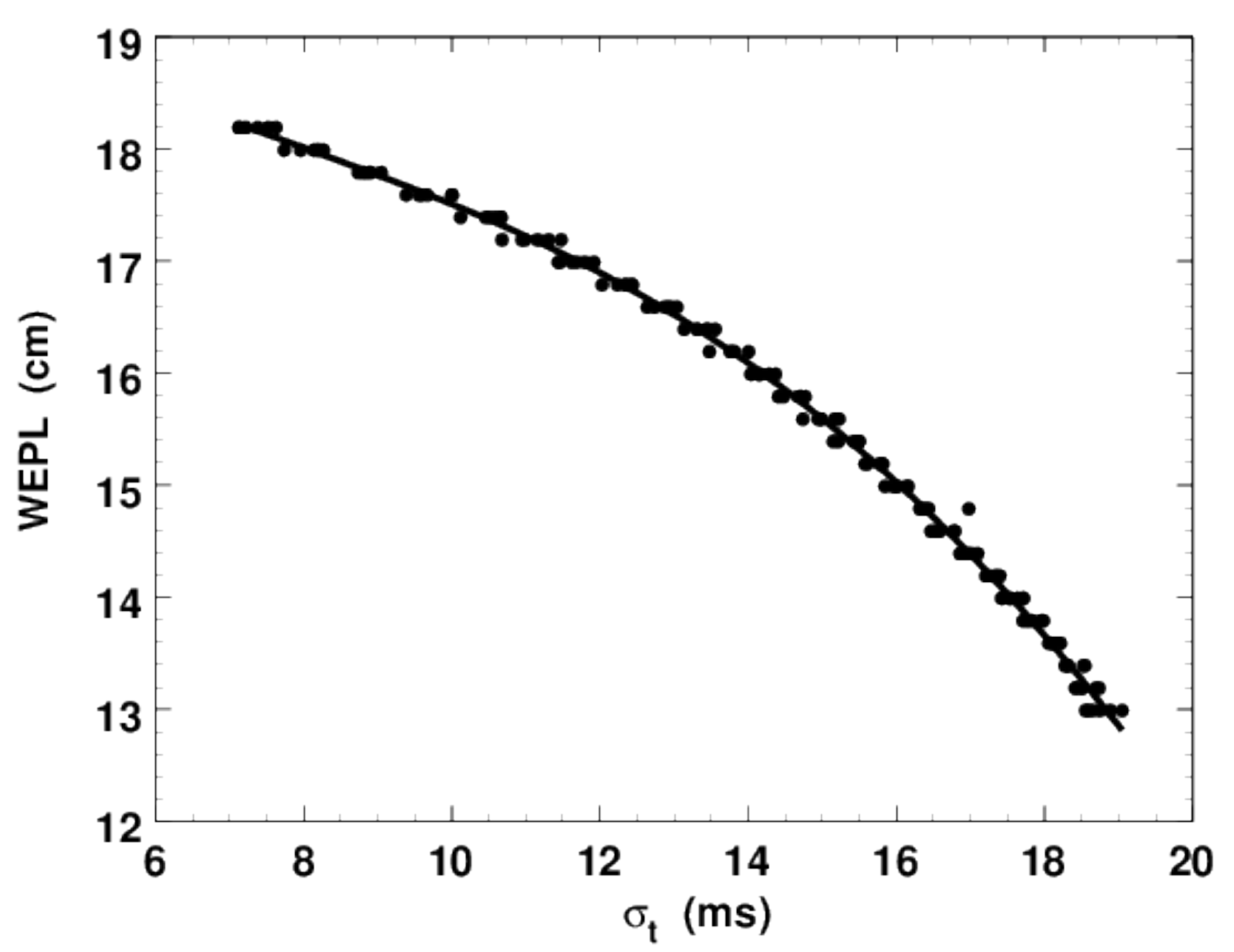} 
\caption{Dots: measured WEPL v. $\sigma_t$ for six DFLR1600 diodes. Line: 5-term (fourth degree) polynomial fit.\label{fig:WEPLfit}}
\end{figure}
\begin{figure}[p]
\centering\includegraphics[width=4.55in,height=3.5in]{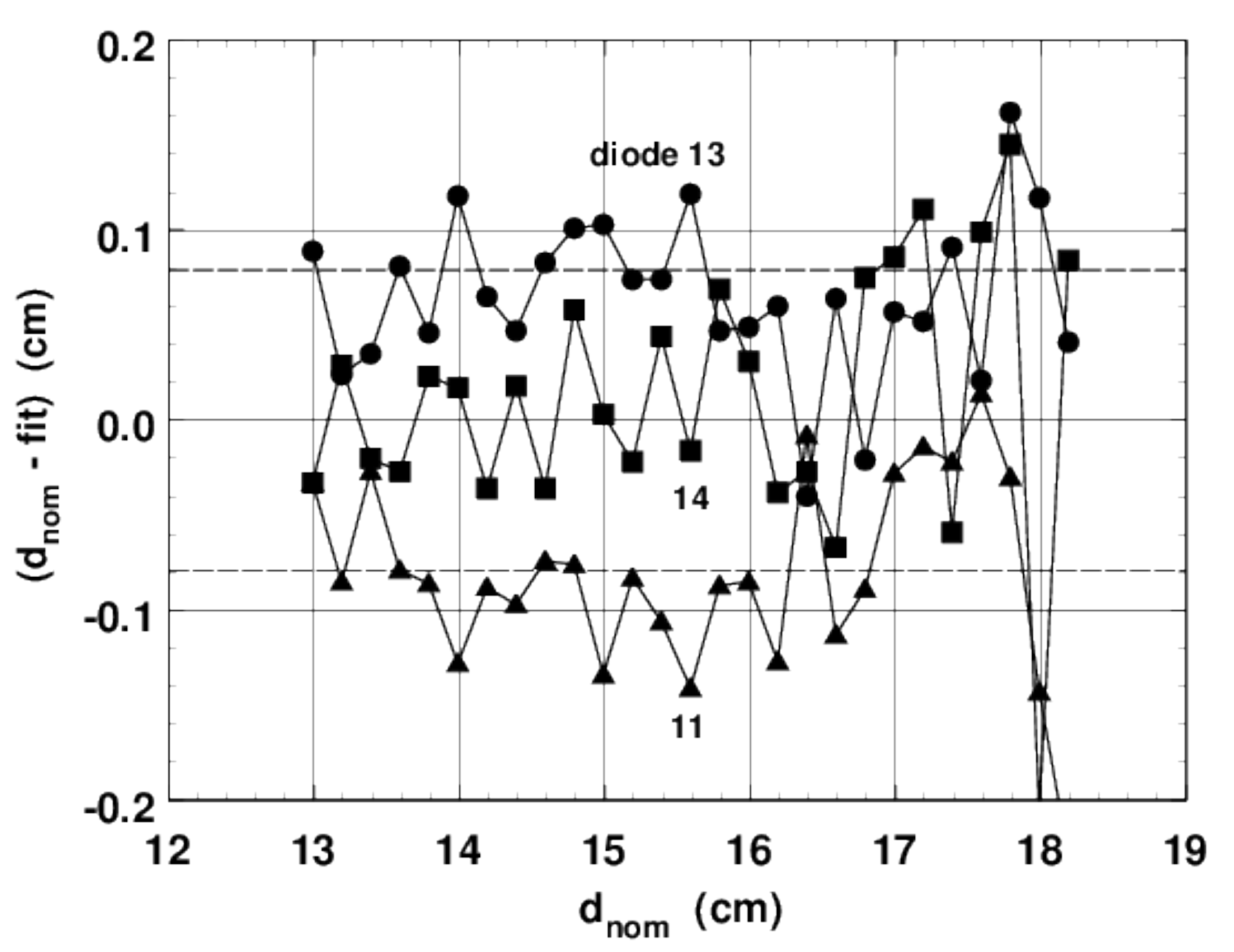} 
\caption{Fit residuals for three diodes. Dashed line: rms residual of global fit.\label{fig:WEPLres}}
\end{figure}
\begin{figure}[p]
\centering\includegraphics[width=4.55in,height=3.5in]{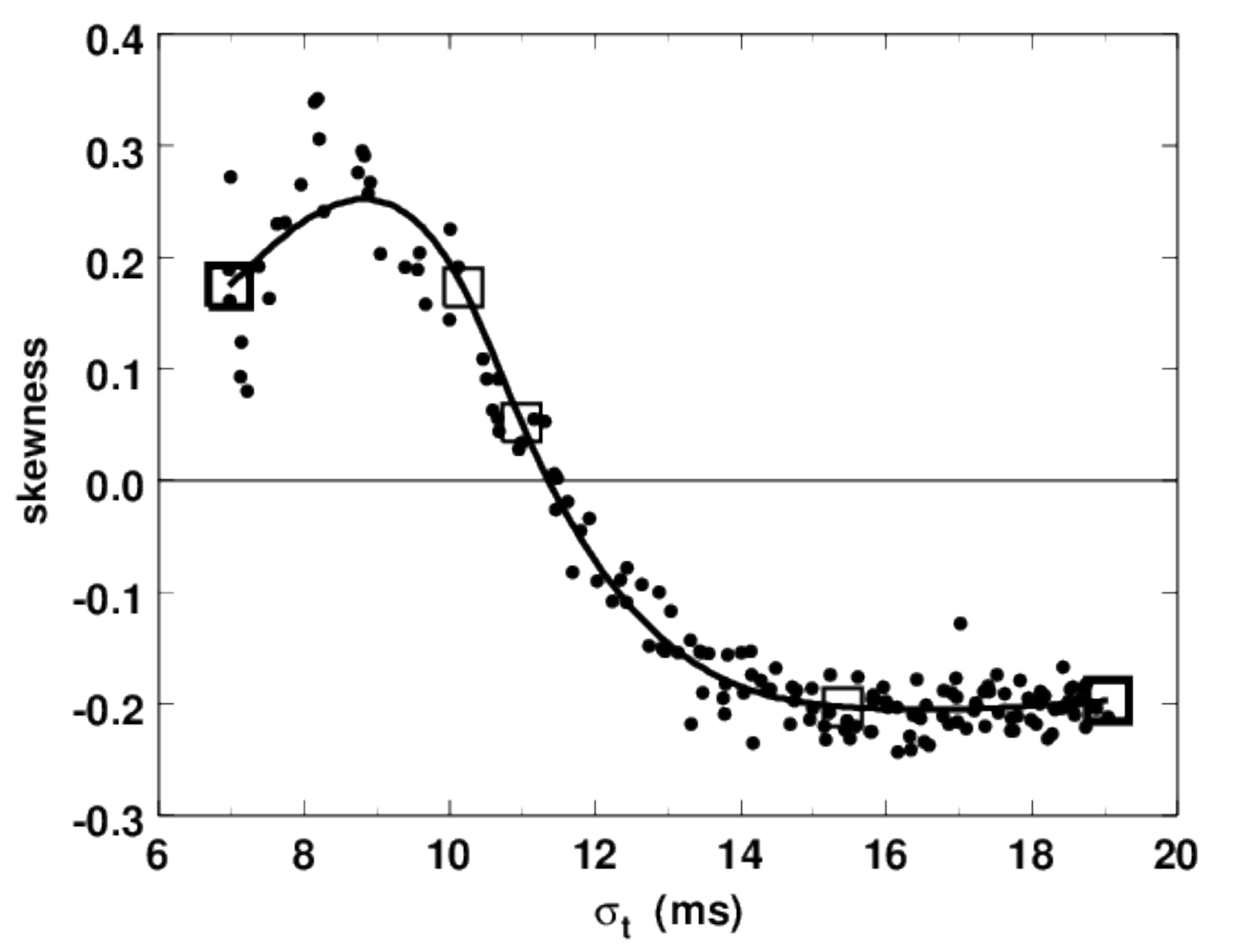} 
\caption{Line: cubic spline fit to natural skewness function. Dots: data. Open squares: spline points (adjustable parameters of the fit). Bold squares: end points.\label{fig:natSkewFit}}
\end{figure}
\begin{figure}[p]
\centering\includegraphics[width=4.55in,height=3.5in]{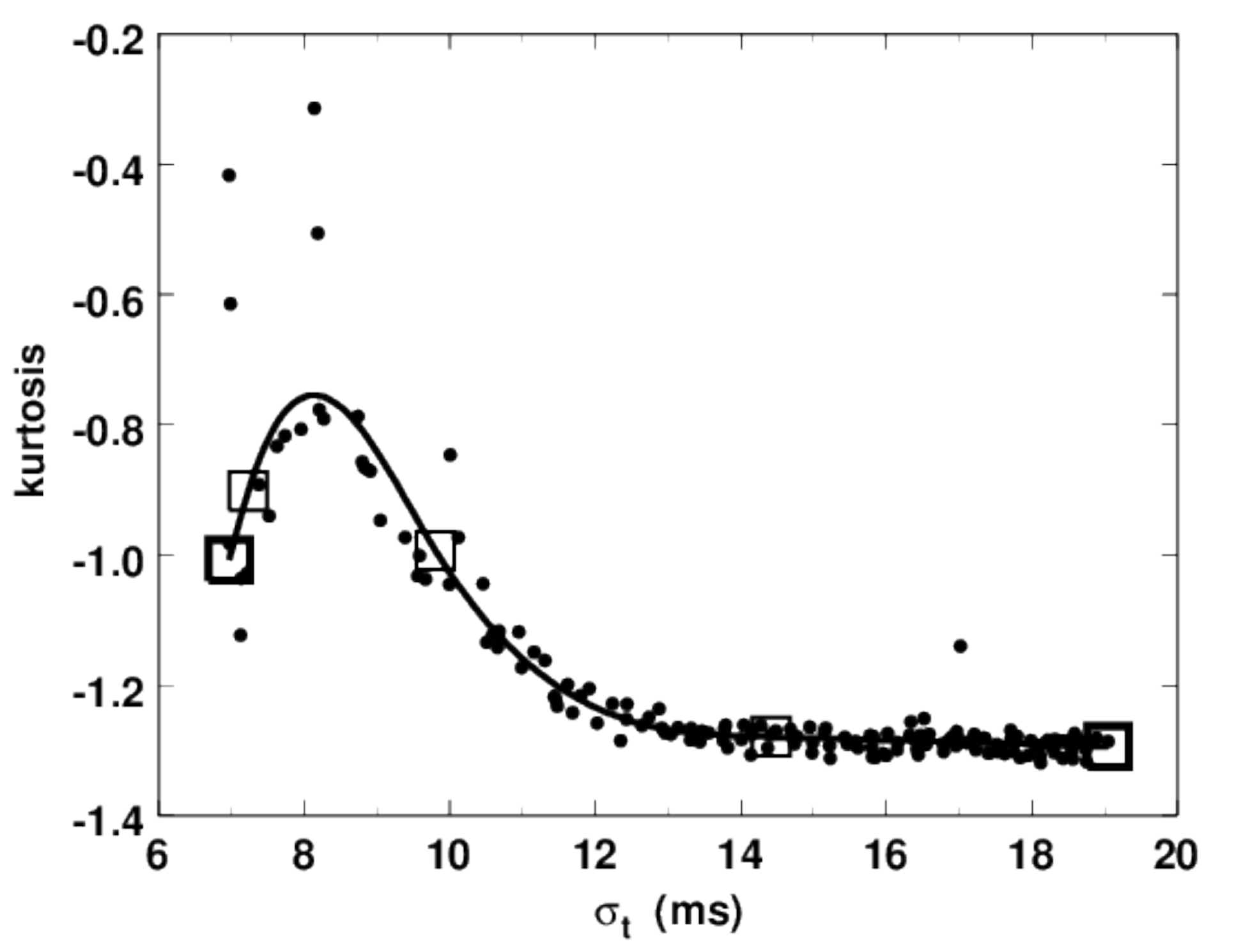} 
\caption{Line: cubic spline fit to natural kurtosis function. Dots: data. Open squares: spline points (adjustable parameters of the fit). Bold squares: end points.\label{fig:natKurtFit}}
\end{figure}
\clearpage

\begin{figure}[p]
\centering\includegraphics[width=4.55in,height=3.5in]{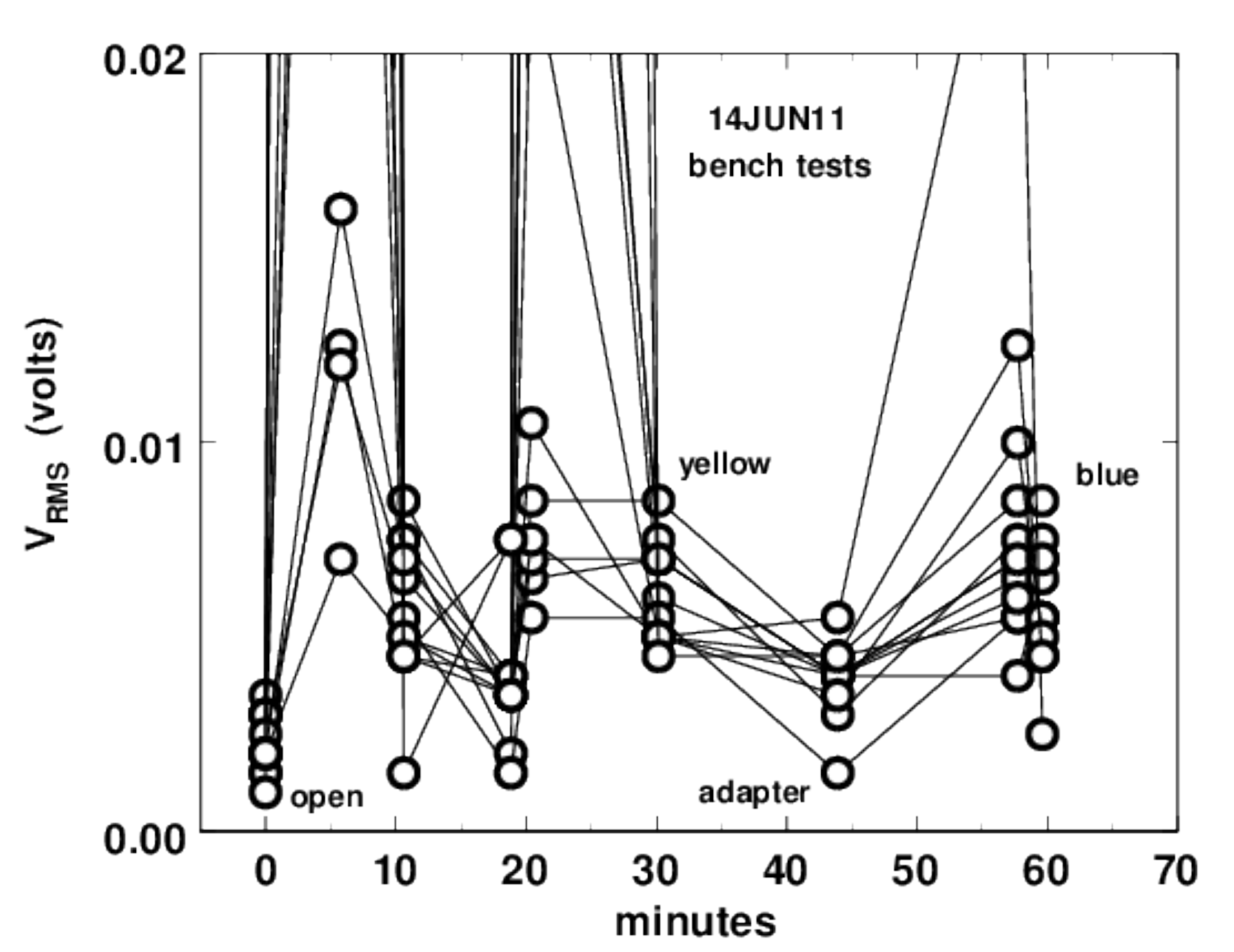} 
\caption{Noise measurements for various cable and adapter configurations.\label{fig:vRMS14JUN11}}
\end{figure}
\begin{figure}[p]
\centering\includegraphics[width=4.55in,height=3.5in]{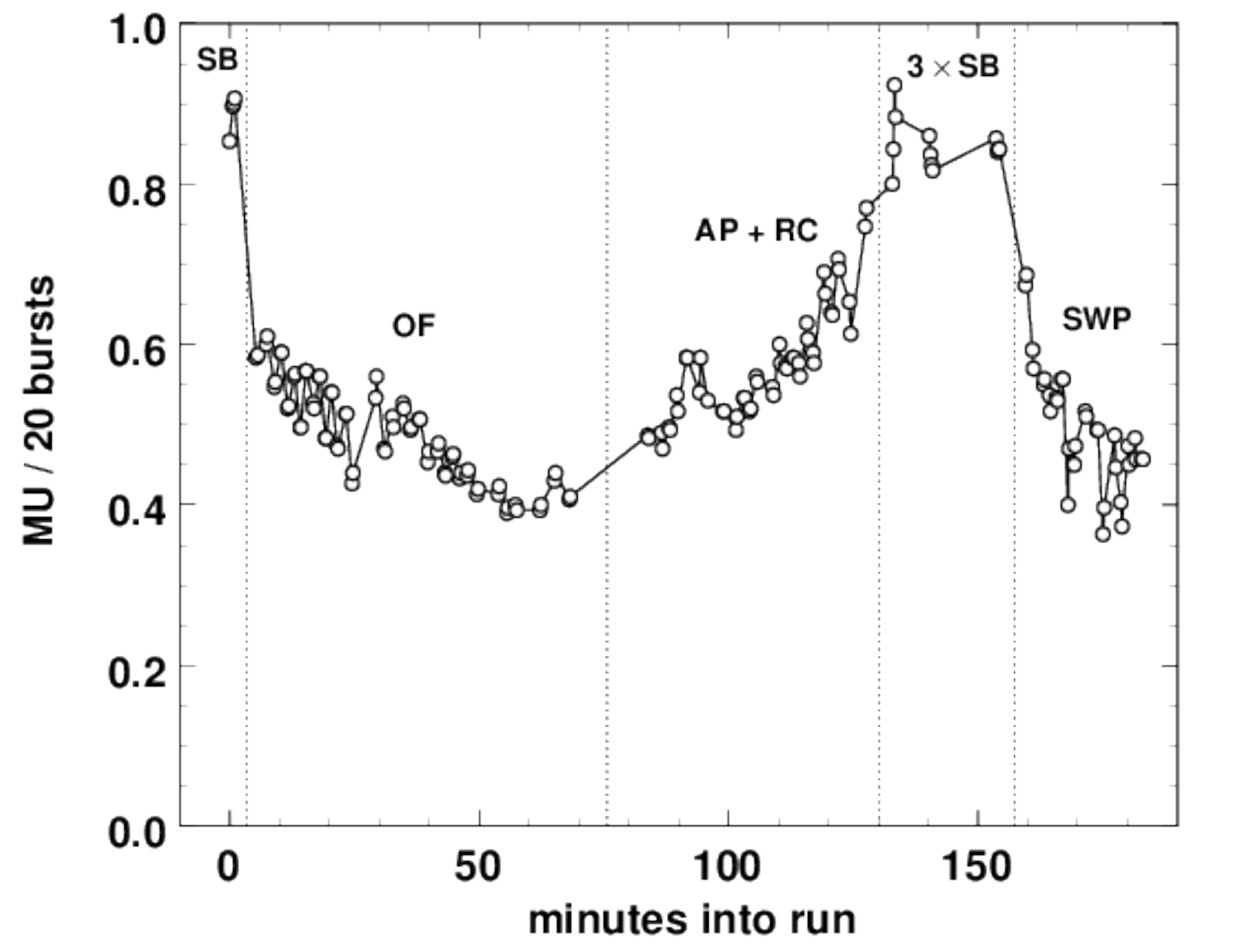} 
\caption{Monitor units (MU) per 20 bursts (modulator cycles) as the 15JUN11 run progressed.\label{fig:MU}}
\end{figure}
\begin{figure}[p]
\centering\includegraphics[width=4.44in,height=3.5in]{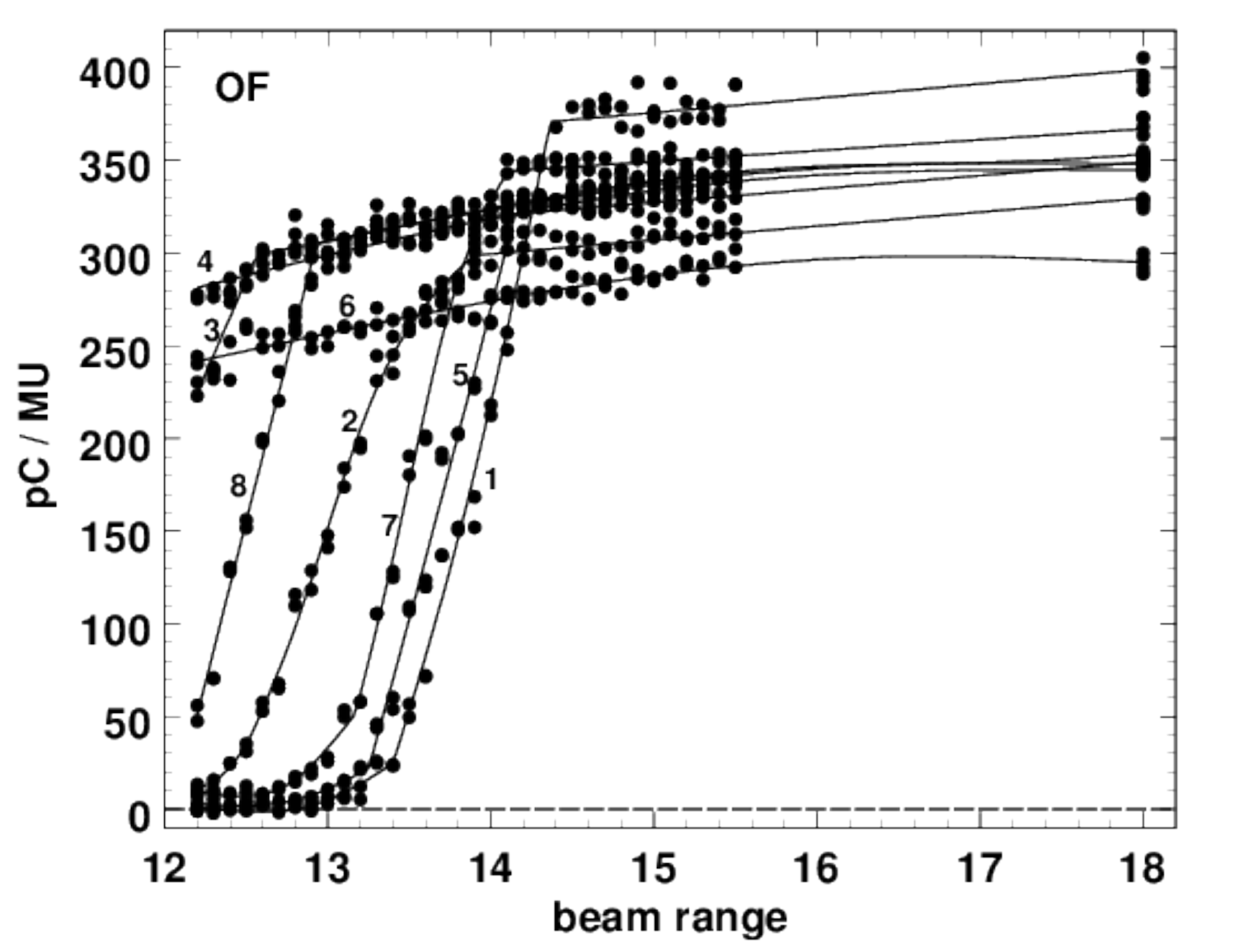} 
\caption{DE results for the Open Field (OF) configuration.\label{fig:extinction1}}
\end{figure}
\begin{figure}[p]
\centering\includegraphics[width=4.44in,height=3.5in]{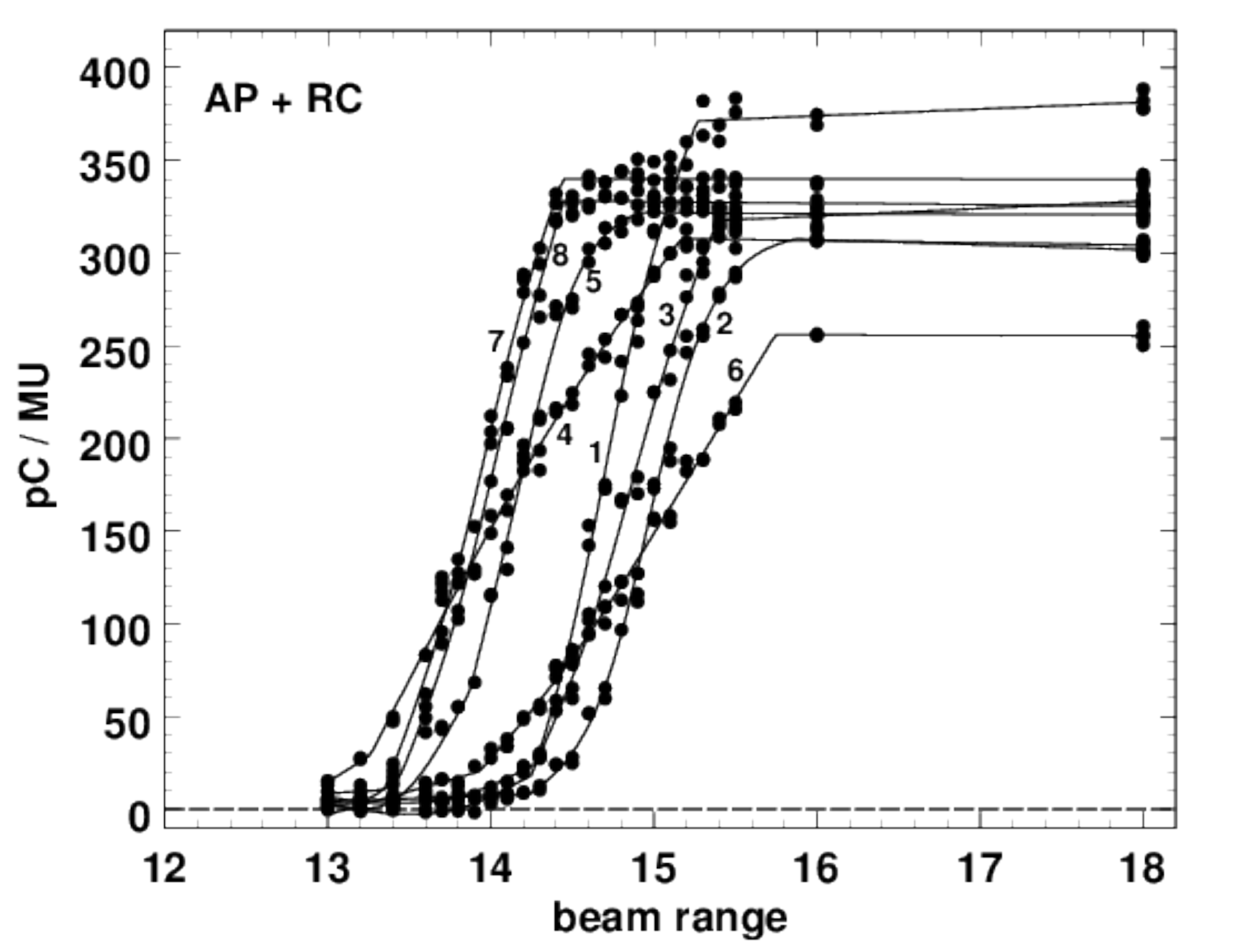} 
\caption{DE results for the Aperture plus Range Compensator (AP + RC) configuration.\label{fig:extinction2}}
\end{figure}
\begin{figure}[p]
\centering\includegraphics[width=4.44in,height=3.5in]{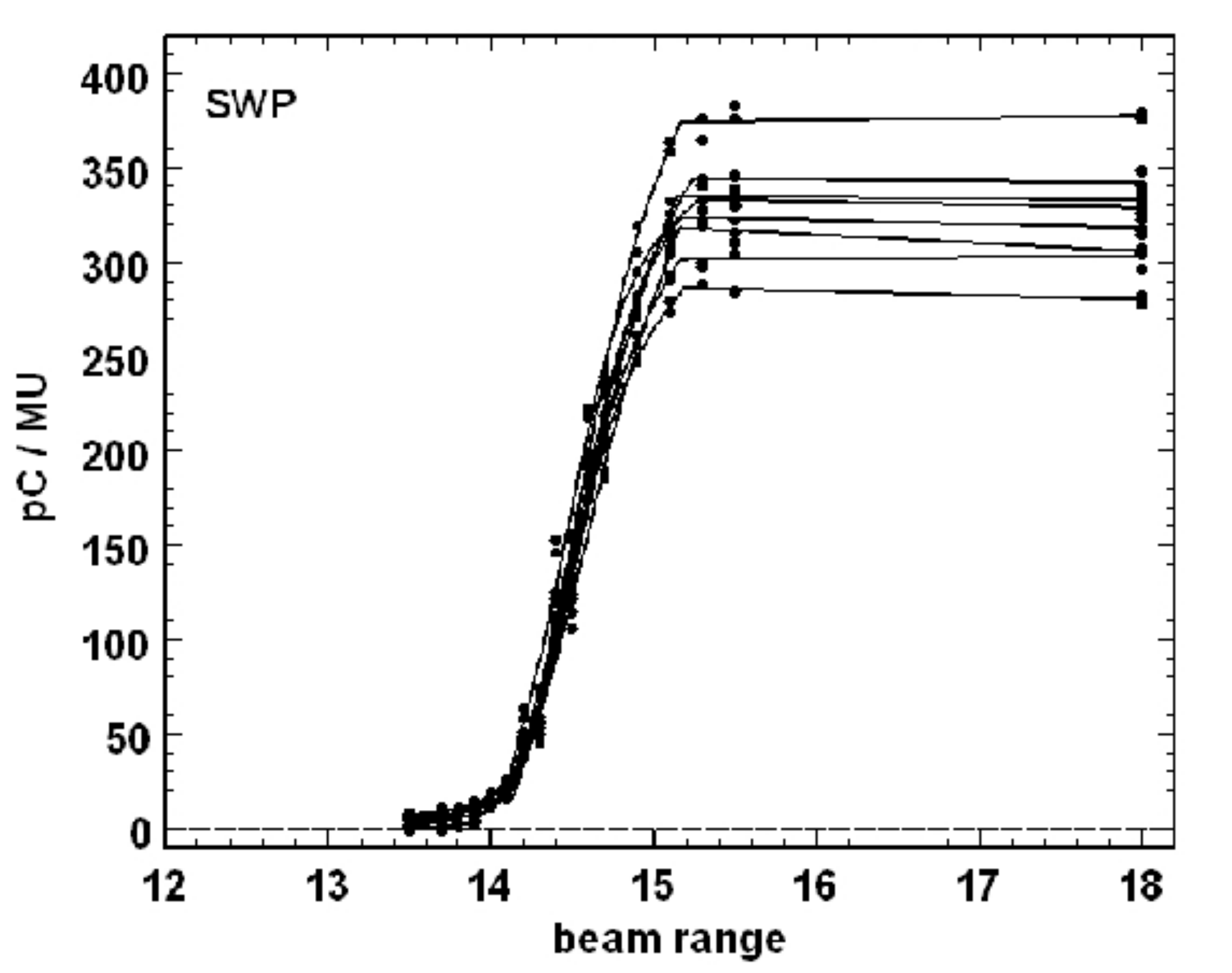} 
\caption{DE results for the Solid Water Phantom (SWP).\label{fig:extinction3}}
\end{figure}
\begin{figure}[p]
\centering\includegraphics[width=4.55in,height=3.5in]{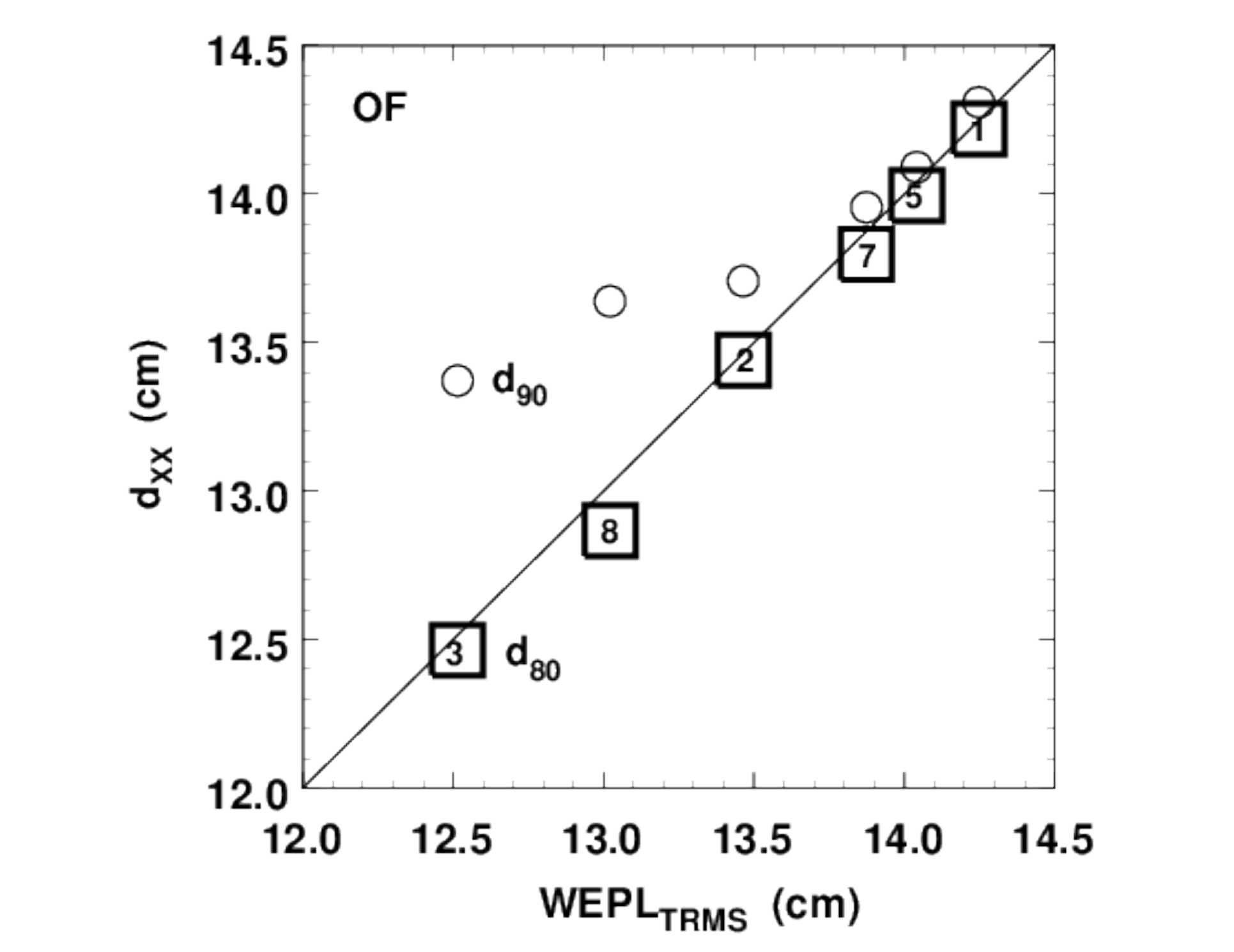} 
\caption{Correlation between WEPL$_\mathrm{TRMS}$ and $d_{80}$ (squares) or $d_{90}$ (circles) for OF configuration.\label{fig:dxxVwepl1}}
\end{figure}
\begin{figure}[p]
\centering\includegraphics[width=4.55in,height=3.5in]{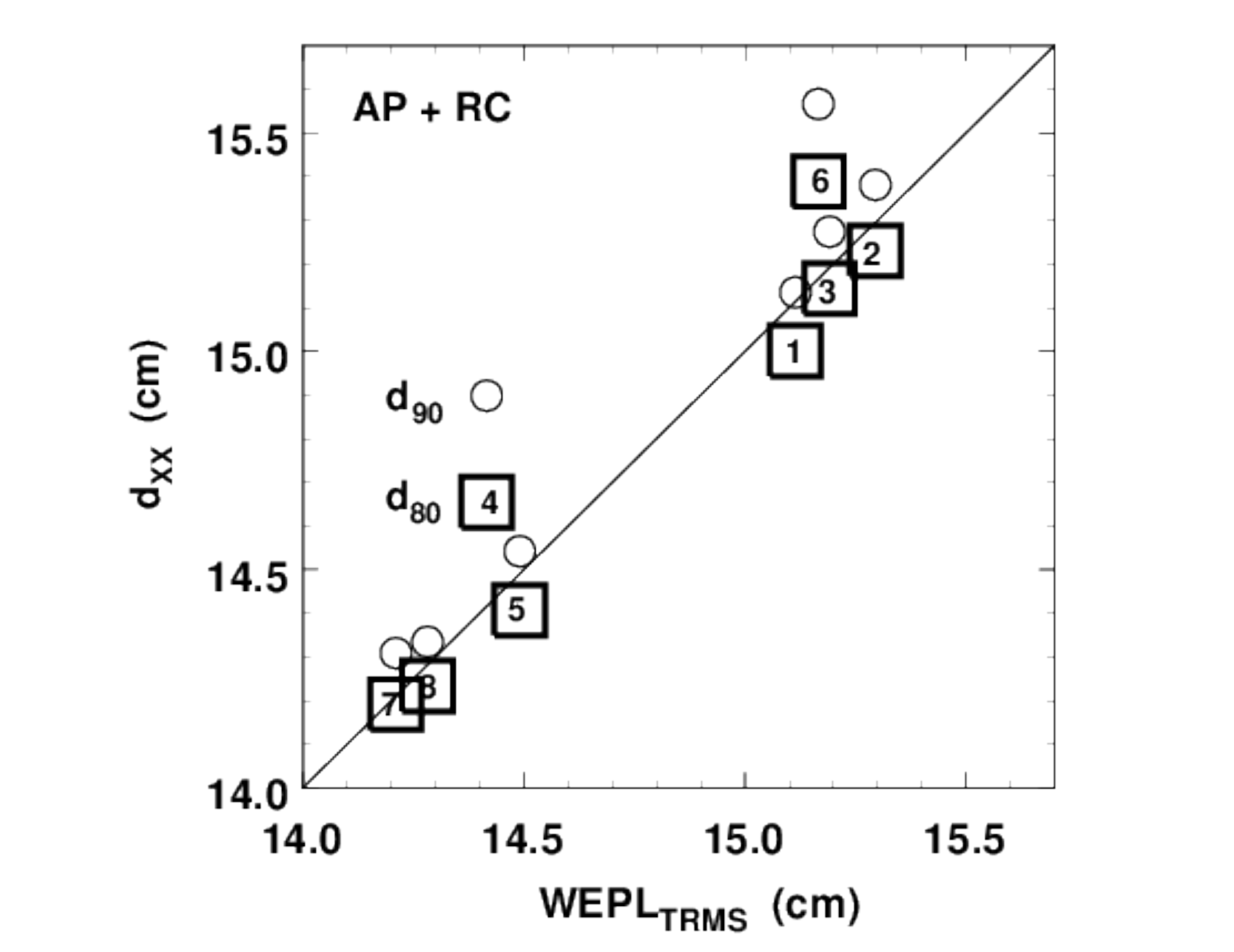} 
\caption{Correlation between WEPL$_\mathrm{TRMS}$ and $d_{80}$ (squares) or $d_{90}$ (circles) for (AP + RC).\label{fig:dxxVwepl2}}
\end{figure}
\begin{figure}[p]
\centering\includegraphics[width=4.55in,height=3.5in]{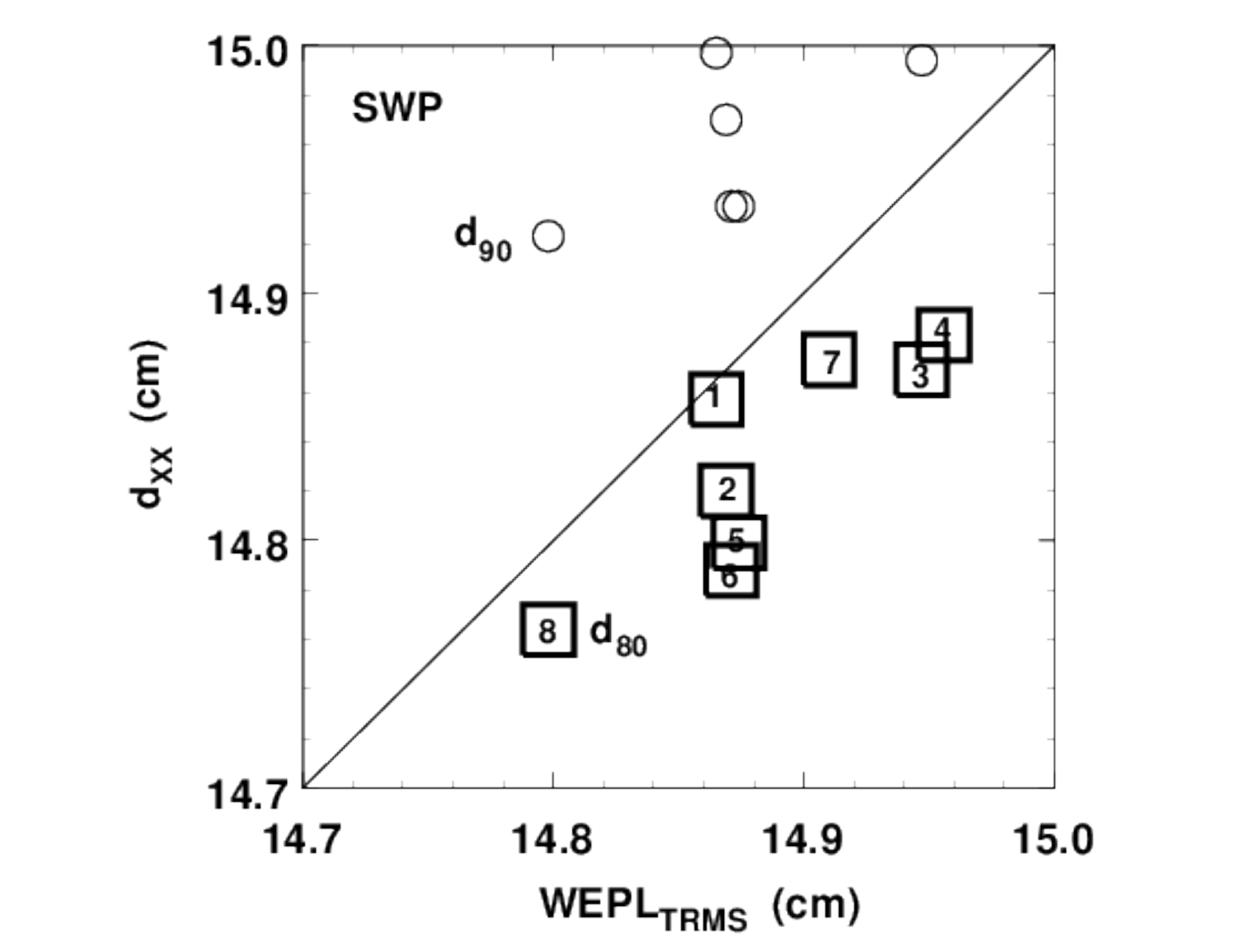} 
\caption{Correlation between WEPL$_\mathrm{TRMS}$ and $d_{80}$ (squares) or $d_{90}$ (circles) for SWP.\label{fig:dxxVwepl3}}
\end{figure}
\begin{figure}[p]
\centering\includegraphics[width=4.44in,height=3.5in]{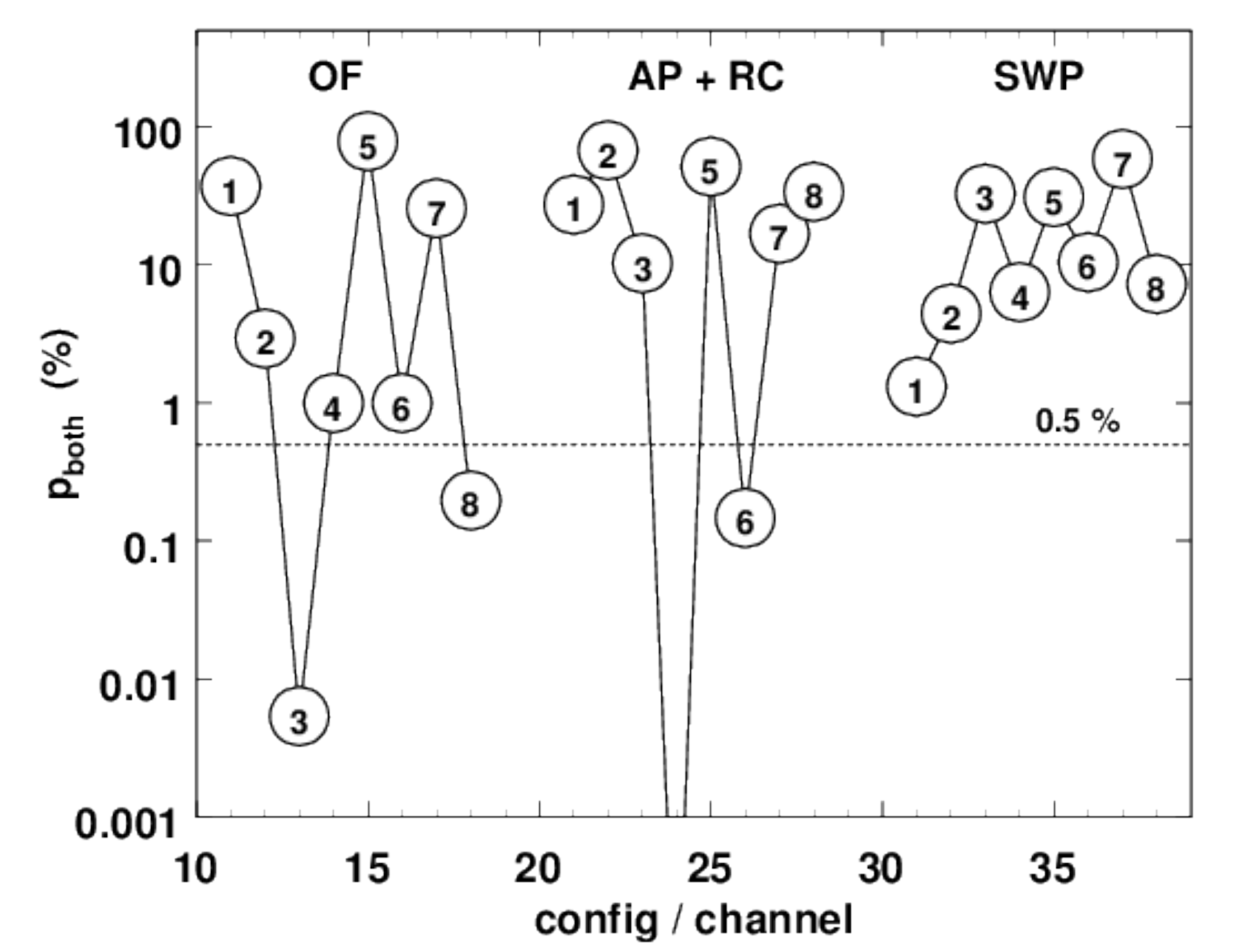} 
\caption{$p_\mathrm{both}$, from the scout beam, for all eight channels and three configurations.\label{fig:pBoth}}
\end{figure}
\clearpage

\end{document}